\documentclass[12pt,twoside,a4paper]{article}
\usepackage{epsfig}
\usepackage{amssymb}
\usepackage{wasysym}
\author{Martin Reuter$^1$ and Jan-Markus Schwindt$^2$}
\date{} 

\title{Scale-dependent metric and causal structures in Quantum Einstein Gravity}

\begin{document}
\maketitle 

\vspace{-8cm}
\begin{flushright} 
MZ-TH/06-22
\end{flushright}
\vspace{6cm}
\centerline{\small\it $^1$ Institute of Physics, University of Mainz, Staudingerweg 7, }
\centerline{\small\it D-55128 Mainz, Germany, E-mail: reuter@thep.physik.uni-mainz.de}
\centerline{\small\it $^2$ Institute of Theoretical Physics, University of Heidelberg, 
  Philosophenweg 16, }
\centerline{\small\it D-69120 Heidelberg, Germany, 
  E-mail: schwindt@thphys.uni-heidelberg.de}

\vspace{0.7cm} 

\begin{abstract}
 Within the asymptotic safety scenario for gravity various conceptual issues related
 to the scale dependence of the metric are analyzed. The running effective field 
 equations implied by the effective average action of Quantum Einstein Gravity (QEG)
 and the resulting families of resolution dependent metrics are discussed.
 The status of scale dependent vs. scale independent diffeomorphisms is clarified,
 and the difference between isometries implemented by scale dependent and independent
 Killing vectors is explained. A concept of scale dependent causality is proposed
 and illustrated by various simple examples. The possibility of assigning an
 ``intrinsic length" to objects in a QEG spacetime is also discussed. 
\end{abstract}

\newpage
\noindent{\bf\large 1. Introduction}

During the past decade a lot of efforts went into the exploration of the nonperturbative
renormalization behavior of Quantum Einstein Gravity \cite{mr}-\cite{hier}. In
\cite{mr} a functional renormalization group (RG) equation for gravity has been
introduced; it defines a Wilsonian RG flow on the theory space consisting of
all diffeomorphism invariant action functionals for the metric 
$g_{\mu\nu}$. In \cite{mr} it has been 
applied to the Einstein-Hilbert approximation which allows for an approximate
calculation of the beta-functions of Newton's constant and the cosmological constant.
The complete flow pattern was found in \cite{frank1}, and higher derivative
truncations were analyzed in \cite{oliver1,oliver2,codello}. Matter fields were
added in refs.~\cite{percadou,perper1}, and in \cite{litimgrav} the beta-functions
of \cite{mr} and \cite{oliver1} were used for finding optimized RG flows.
The most remarkable result of these investigations is that the beta-functions
of \cite{mr} predict a non-Gaussian RG fixed point \cite{souma}. After detailed
studies of the reliability of the pertinent truncations 
\cite{oliver1,frank1,oliver2,litimgrav} it is now believed 
that it corresponds to a fixed point in the exact theory and is not an approximation 
artifact. It was found to possess all the necessary properties to make quantum 
gravity nonperturbatively renormalizable along the lines of Weinberg's ``asymptotic
safety" scenario \cite{wein,livrev}, thus overcoming the notorious problems
related to its nonrenormalizability in perturbation theory. We shall refer to the 
quantum field theory of the metric tensor whose infinite cutoff limit is taken at 
the non-Gaussian fixed point as Quantum Einstein Gravity or ``QEG". This theory
should not be thought of as a quantization of classical general relativity.
Its bare action is dictated by the fixed point condition and is therefore expected
to contain more invariants than the Einstein-Hilbert term only. Independent evidence
pointing towards a fixed point in the full theory came from the symmetry reduction
approach of ref.~\cite{max} where the 2-Killing subsector of the gravitational path 
integral was quantized exactly.

Except for the latter investigations, all recent studies of the asymptotic safety 
scenario in gravity made use of the approach outlined in \cite{mr}. It is based 
upon the concept of the effective average action \cite{avact,ym,avactrev}, a specific
continuum implementation of the Wilsonian renormalization group. In its original form 
for matter theories in flat spacetime it has been applied to a wide
range of problems both in particle and statistical physics. As compared to
alternative functional RG approaches in the continuum \cite{bagber} the average 
action has various crucial advantages; the most important one is its similarity with 
the standard effective action $\Gamma$. In fact, the average action is a scale dependent 
functional $\Gamma_k$ depending on a ``coarse graining" scale $k$ which approaches 
$\Gamma$ in the limit $k \rightarrow 0$ and the bare action $S$ in the limit 
$k \rightarrow\infty$. The close relationship of $\Gamma_k$ and the standard $\Gamma$
was often crucial for finding the right truncations of theory space encapsulating the 
essential physics. From conventional field theory we have a well-trained intuition
about what a typical effective action $\Gamma$ should look like, and we can now use this
experience in order to guess, and subsequently verify (or falsify) by explicit 
computations what the important terms in $\Gamma_k$ are. For the functionals evolved
by the older exact RG equations a comparable understanding is lacking ususally.

Another advantage of the average action is that it defines a family of effective field 
theories $\{ \Gamma_k, 0 \leq k < \infty \}$ labeled by the coarse graining scale $k$.
If a physical situation involves only a single mass scale, then it is well described
by a tree level evaluation of $\Gamma_k$, with $k$ chosen to equal that scale.
In particular, the stationary points of $\Gamma_k$ have the interpretation of a 
$k$-dependent field average (approaching the standard 1-point function for 
$k \rightarrow 0$). The quality of the 
effective field theory description depends on the size of the
fluctuations relative to the average field.

In gravity the effective average action of \cite{mr} is a diffeomorphism invariant
functional of the metric: $\Gamma_k[g_{\mu\nu}]$. Here the analogous average field 
$\langle g_{\mu\nu}\rangle_k$ satisfies the ``effective Einstein equations"
\begin{equation}\label{fe}
 \frac{\delta\Gamma _k}{\delta g_{\mu\nu}(x)}\left [ \left < g 
 \right >_k \right ]=0.
\end{equation} 
A given quantum state $| \Psi \rangle$ of the gravitational field implies an infinite
family of average metrics: $\{ \langle g_{\mu\nu}\rangle_k,\; 0 \leq k < \infty \}$.
A scale dependence of the metric \cite{nino} has profound consequences since
$\langle g_{\mu\nu}\rangle_k$ describes a geometry of spacetime which depends on 
the degree of ``coarse graining", or the ``resolving power" of the ``microscope"
with which
it is looked upon. In the case of QEG, it has been shown \cite{oliver1,oliver2}
that this scale dependence leads to fractal properties of spacetime, and that in
the scaling regime of the non-Gaussian fixed point, corresponding to sub-Planckian
distances, the fractal dimension of spacetime equals 2. In particular, making 
essential use of (\ref{fe}) and the effective field theory properties of $\Gamma_k$,
the spectral dimension \cite{ajl1} has been calculated; it was found to interpolate
between 4 at macroscopic, and 2 at microscopic distances \cite{oliverfrac}.
In \cite{ncgeom}, Connes et al. speculated about the possible relevance of this
dimensional reduction for the noncommutative geometry of the standard model.
Remarkably, exactly the same dimensional reduction has been found in Monte Carlo
simulations within the causal dynamical triangulation approach \cite{ajl1,ajl2,ajl34}.

The purpose of the present paper is to discuss in detail the conceptual status of
the metric families $\{ \langle g_{\mu\nu}\rangle_k,\; 0 \leq k < \infty \}$
and to illustrate, by means of simple examples, the novel physical effects which arise
from a $k$-dependence of the spacetime geometry. In particular we argue that there
is a well-defined notion of a scale dependent causality. We also
analyze the question how, and to what extent, 
geometric structures or material objects in a QEG
spacetime can be ascribed an ``intrinsic" length which one would then consider 
``the" length of the objects.

The motivation for this work is that one would like to extract as much physical 
information as possible directly from the RG flow. Up to now this was mostly
done by some form of ``renormalization group improvement" \cite{bh}-\cite{mof}
whereby $k$ is identified with some scale typical for the physical situation
under study. The notorious difficulty of this method consists in finding the correct
``cutoff identification". Moreover, even if by some high degree of symmetry, for
instance, this identification is uncontroversial, the disadvantage is that only spacetime
properties on a single typical scale
are described, albeit on a dynamically natural one. In the present
paper, the idea is to completely abandon the ``cutoff identification". Instead we
try to ``visualize" the stock of Riemannian structures $\{ \langle g_{\mu\nu}\rangle_k \}$
as a whole and to deduce information about the physical properties of the QEG spacetimes
from it.

Before closing this introduction, let us be slightly more explicit about the construction 
of $\Gamma_k$ for gravity \cite{mr}. Its (formal) starting point
is the path integral $\int {\cal D}\gamma _{\mu\nu}
\exp \left ( -S[\gamma]\right )$ over all metrics $\gamma _{\mu\nu}$, gauge fixed by
means of a background gauge fixing condition \cite{back}. Even without an infrared cutoff,
upon introducing sources and performing a Legendre 
transformation one is led to an effective action $\Gamma\left [ g_{\mu\nu};
\bar{g}_{\mu\nu}\right ]$ which depends on two metrics, the expectation value of 
$\gamma _{\mu\nu}$, denoted $g_{\mu\nu}$, and the non-dynamical background field
$\bar{g}_{\mu\nu}$. It is well-known \cite{back} that the functional 
$\Gamma[g_{\mu\nu}]\equiv\Gamma[g_{\mu\nu};\bar{g}_{\mu\nu}=g_{\mu\nu}]$ 
obtained by equating the two
metrics generates a possible set of 1PI Green's functions of the theory.
The average action has a built-in, variable IR cutoff. This
IR cutoff is implemented by first expanding the shifted integration variable
$h_{\mu\nu}\equiv\gamma _{\mu\nu}-\bar{g}_{\mu\nu}$ 
in terms of eigenmodes of $\bar{D}^2$,
the covariant Laplacian formed with the background metric 
$\bar{g}_{\mu\nu}$, and interpreting ${\cal D}h_{\mu\nu}$ as an
integration over all expansion coefficients.
Then a suppression term is introduced which damps the
contribution of all $\bar{D}^2$-modes with eigenvalues smaller than $k^2$.
Following the usual steps \cite{avactrev,bagber} this leads to the scale dependent
functional $\Gamma_k[g_{\mu\nu};\bar{g}_{\mu\nu}]$, and again the action with one
argument is obtained by equating the two metrics:
$\Gamma_k[g_{\mu\nu}]\equiv\Gamma_k[g_{\mu\nu};\bar{g}_{\mu\nu}=g_{\mu\nu}]$.
It is this action which appears in (\ref{fe}). 

At least when one applies the average action technique to Euclidean non-gauge 
theories on flat space,
$\Gamma _k$ may be interpreted as arising from a continuum version of a
Kadanoff-Wilson block spin procedure, i.e. it defines the dynamics of ``coarse
grained" dynamical variables which are averaged over a certain region of
Euclidean spacetime. Denoting the typical linear extension of the averaging region 
by $\ell$, one has $\ell \approx \pi /k$ in flat spacetime.
In this sense, $\Gamma _k$ can be thought of as a ``microscope" with an
adjustable resolving power $\ell =\ell(k)$.
In quantum gravity where the metric is dynamical 
the relationship between the IR
cutoff $k$ and the averaging scale $\ell$ is more complicated in general.
We will return to this issue in section 3.

The running action $\Gamma_k$ satisfies an
exact functional RG equation \cite{mr}. In practice it is usually solved on a
truncated theory space. In the Einstein-Hilbert truncation
of pure gravity $\Gamma_k$ is approximated by a functional of the form
\begin{eqnarray}
\label{3in2}
\Gamma_k[g]=\left(16\pi G(k)\right)^{-1}\int d^4x\,\sqrt{g}\left\{
-R(g)+2 \Lambda (k)\right\}
\end{eqnarray}
involving a running Newton constant $G(k)$ and cosmological constant 
$\Lambda (k)$.
For each $k$, the action (\ref{3in2}) implies an effective field equation
which happens to be of the form of the classical Einstein equation:
\begin{equation}\label{fieldeq}
 R_{\mu\nu}(\left < g \right > _k)=\Lambda (k) \left < g_{\mu\nu}\right > _k.
\end{equation}
Note that the running Newton constant $G(k)$ does not appear in this effective
Einstein equation. It enters only when matter fields are introduced. In this case
it reads
\begin{equation}\label{m1}
 G_{\mu\nu}(\left < g \right >_k)=-\Lambda(k)\left < g_{\mu\nu}\right >_k
 +8 \pi G(k) \left < T_{\mu\nu}\right >_k   
\end{equation}  
where the scale dependent energy momentum tensor is given by the functional derivative
of the matter part of the average action, $\Gamma _k^{\rm M}[g_{\mu\nu},\chi]$:
\begin{equation}\label{m2}
 \left < T^{\mu\nu}(x)\right >_k \equiv \frac{2}{\sqrt{-g}}\frac{\delta\Gamma_k^{\rm M}
 \left [ \left < g \right >_k, \left < \chi \right >_k \right ]}{\delta g_{\mu\nu}(x)}.   
\end{equation}
Eq.~(\ref{m1}) is coupled to the equation of motion
\begin{equation}\label{m2b}
 \frac{\delta\Gamma_k^{\rm M}}{\delta\chi(x)}
 \left [ \left < g \right >_k, \left < \chi \right >_k \right ]=0.    
\end{equation}
Here $\chi$ denotes the set of matter fields, and 
$\left\{\left < g \right >_k , \left < \chi \right >_k \right\}$ 
is a solution to the coupled gravity plus matter field
equations. The effective Einstein equations (\ref{m1}) are consistent since,
for $\chi$ ``on shell", the energy momentum tensor is covariantly conserved:
\begin{equation}\label{m2a}
 D_\mu (\left < g \right >_k)\left < T^{\mu\nu}(x)\right >_k =0.   
\end{equation}
Here the connection which defines the covariant derivative is built from 
$\left < g \right >_k$. Eq.~(\ref{m2a}) is a consequence of the diffeomorphism invariance
of $\Gamma _k^{\rm M}$.

As for the RG trajectories following from the Einstein-Hilbert
approximation \cite{frank1},
there are several types of them, conveniently plotted in terms of the
dimensionless parameters 
$g(k) \equiv k^2 G(k)$ and $\lambda (k) \equiv \Lambda (k) /k^2$. Among them,
``Type IIIa" trajectories are the type that is 
presumably realized in the real universe since it is the only type that has a 
positive Newton's constant $G(k)$ and a small positive cosmological constant 
$\Lambda(k)$ at macroscopic scales. 
The Type IIIa trajectory contains the following four parts, with
increasing values of the cutoff $k$:  \\ 
i) The classical regime for small $k$ where the trajectory 
is identical to a canonical one, with $G=$const, $\Lambda =$const.\\
ii) The turnover regime where the trajectory, close to the Gaussian
fixed point at $g=\lambda =0$, begins to depart from the canonical
one and turns over to the separatrix which connects the Gaussian 
with the non-Gaussian fixed point $(g_*,\lambda _*)$. By definition, the 
coordinates of the turning point $T$ are $g_T$ and $\lambda _T$, and it is passed 
at the scale $k=k_T$.\\
iii) The growing $\Lambda$ regime where $G(k)$ is approximately constant but 
$\Lambda(k)$ runs proportional to $k^4$. \\
iv) The fixed point regime where the trajectory approaches the non-Gaussian fixed
point in an oscillating manner. Directly at the fixed point one has $g(k)\equiv g_*$
and $\lambda(k)\equiv\lambda _*$, and therefore $G(k) \propto k^{-2}$ and
$\Lambda (k) \propto k^2$ for $k \rightarrow\infty$. The non-Gaussian fixed point
is responsible for the nonperturbative renormalizability of the theory.\\ 
The behavior of the trajectory in the extreme IR is not yet known since the
Einstein-Hilbert approximation breaks down when $\lambda(k)$ approaches 
1/2. A more general truncation is needed to approximate the RG trajectory
in that region. For this reason the classical region i) does not necessarily
extend to $k=0$, and we speak about ``laboratory" scales for values of 
$k \equiv k_{\rm lab}$ in the region where $G$ and $\Lambda$ are constant.

The remaining sections of this paper are organized as follows. In section 2 we discuss 
various conceptual issues related to the families $\{ \langle g_{\mu\nu}\rangle_k \}$,
in particular their connection with the quantum state, the status of $k$-independent vs.
$k$-dependent diffeomorphisms, and symmetries of QEG spacetimes. In this section we also
explain the idea of a scale dependent causality. Section 3 is devoted to ``$k$-microscopes",
a universal and mathematically simple model of an experimental setup for the observation of
the spacetime structure. The sections 4 and 5 contain various illustrative examples;
in section 4 a family of Schwarzschild-de Sitter metrics is analyzed, and section 5 deals
with Robertson-Walker families. In section 6 the concept of an ``intrinsic scale" is
described and its viability is tested in several examples. Section 7 contains a summary
of the results.

At this point the reader should be warned already that the families of metrics considered
in the examples do not yet correspond to realistic measurements or observations. In order
to be able to find analytic solutions $ \langle g_{\mu\nu}\rangle_k$ we usually require
the spacetimes to be highly symmetric, even in presence of the ``microscope".
Sometimes this has the effect of overdrawing the novel effects due to the $k$-dependence
of the metric so that they might appear somewhat ``exotic". (In the black hole examples,
for instance, the symmetry requirements amount to the assumption of a ``microscope" 
which is much larger than the black hole itself.) As we are mostly interested in matters
of principle here we shall not try to be very realistic in this respect.\\

\noindent{\bf\large 2. QEG spacetimes}

In the following we assume 
that we have solved the exact RG equation and picked a specific RG 
trajectory emanating from the non-Gaussian fixed point. This trajectory completely 
defines the quantum theory of the gravitational field then, in the sense that 
all free parameters characterizing the RG trajectories are given fixed values. In
a standard field theory such as QED, say, this fixing of parameters corresponds to 
identifying the renormalized values of the electron's mass and charge with their measured
values.\\

\noindent{\bf 2.1. State dependence}

Leaving technical issues aside for the moment \cite{livrev} it should be possible to 
reformulate the theory resulting from a given trajectory in a Hilbert space language.
In particular, one should be able to interpret the correlators of the path integral 
approach, $\int\mathcal{D}\gamma \;\gamma_{\mu_1 \nu_1}(x_1)...\gamma_{\mu_n \nu_n}(x_n)
\exp (-S[\gamma])$, as expectation values involving the metric operator
$\hat{g}_{\mu\nu}(x)$ and a certain state $|\Psi\rangle$:
\begin{equation}\label{corr}
 \langle\Psi | \hat{g}_{\mu_1 \nu_1}(x_1)... \hat{g}_{\mu_n \nu_n}(x_n)| \Psi\rangle .   
\end{equation}
Within the path integral formalism, the dependence on the state $|\Psi\rangle$ is 
encoded in the boundary conditions imposed on the fields integrated over. While this
state dependence is of course central in the path integral approach applied to the 
elementary quantum mechanics of point particles , its importance is deemphasized in the 
standard matter field theories on a nondynamical Minkowski space. In the latter case
$|\Psi\rangle$ is usually taken to be the essentially unique 
Poincar$\acute{\rm e}$ invariant vacuum 
state. As there is no a priori distinguished (vacuum) state in quantum gravity we shall
not try to fix $|\Psi\rangle$ here and to relate it to the boundary conditions for the 
path integral. However, from a conceptual point of view it  
will be important to keep in mind
that in principle correlators such as (\ref{corr}) do depend on the state of the 
gravitational field, and that the corresponding path integral incapsulates this state
dependence via boundary conditions and/or surface terms in the action (see \cite{livrev}
for a more detailed discussion.)

The effective average action is defined in terms of the path integral
$\int\mathcal{D}\gamma_{\mu \nu}\exp (-S[\gamma])$ with additional mode suppression and
source terms included. As a result, the functional $\Gamma_k$, too, has an implicit 
dependence on the state $|\Psi\rangle$.

This remark applies to the standard effective action $\Gamma=\lim_{k \rightarrow 0}
\Gamma _k$ already. In fact since, by the usual arguments, the expectation value
\begin{equation}\label{vev}
 \langle \Psi | \hat{g}_{\mu\nu}(x) | \Psi\rangle \equiv g_{\mu\nu}(x)   
\end{equation}
is a critical point of $\Gamma$,
\begin{equation}\label{eff}
 \frac{\delta\Gamma}{\delta g_{\mu\nu}}[g]=0,   
\end{equation}
it is clear that $\Gamma$ ``knows" about the choice for $|\Psi\rangle$.
In general the effective field equations (\ref{eff}) will have many more 
solutions than just the expectation value (\ref{vev}). Uniqueness of the solution
could be achieved by imposing subsidiary conditions on $g_{\mu\nu}(x)$. 
In principle it should 
be possible to derive those subsidiary conditions from the path integral over 
$\gamma_{\mu\nu}$.

For $k>0$, eq.~(\ref{eff}) is replaced by the scale dependent effective field equation
(\ref{fe}) and the situation is similar. A given $|\Psi\rangle$ translates to given
boundary conditions for the path integral defining $\Gamma_k$. Then, by the very construction 
of the effective average action, the one-point function $\langle g_{\mu\nu}(x)\rangle_k$
defined by the cut-off path integral $\int\mathcal{D}\gamma \exp(-S[\gamma]-
\Delta_k S+ \cdots)$ is known to be 
a solution of (\ref{fe}). However, in general (\ref{fe}) 
will have many more solutions . To find out which one among them is 
$\langle g_{\mu\nu}(x)\rangle_k$, one should again restrict the space of allowed solutions 
by subsidiary conditions which are to be derived from the path integral and hence ``know"
about the state $|\Psi\rangle$.

The derivation of such subsidiary conditions is a formidable task, well beyond the present
technical state of the art. In this paper we shall therefore not try to impose such conditions 
but rather analyze the space of {\it all} solutions to the effective field equations
(of a given symmetry type), keeping in mind, however, that not all solutions necessarily 
come from a physically acceptable state $|\Psi\rangle$.\\  

\noindent{\bf 2.2. Scale dependent metric structure}

Each state implies a family of mean field metrics
$\left \{ \left < g_{\mu\nu}\right > _k (x) ;\; 0 \leq k< \infty\right \}$, solving
the family of effective Einstein equations (\ref{fe}) along the chosen trajectory.
As for their interpretation, it is important to note 
that the infinitely many equations in (\ref{fe}), one at
each scale $k$, are valid simultaneously, and that all the mean fields 
$\left < g_{\mu\nu} \right >_k$ refer to one and the same physical ``system",
a state $|\Psi \rangle$ of the ``quantum spacetime" in the QEG sense. 
The mean fields $\left < g_{\mu\nu} \right >_k$
describe the metric structure in dependence on the length scale on which the
spacetime manifold is probed. An observer exploring the structure of spacetime
using a ``microscope" of resolution $\ell(k)$ will perceive the universe as a
Riemannian manifold with the metric $\left < g_{\mu\nu} \right >_k$. While
$\left < g_{\mu\nu} \right >_k$ is a smooth classical metric at every fixed $k$,
the quantum spacetime can have fractal properties because on different scales
different metrics apply. In this sense the metric structure on the quantum
spacetime is given by an infinite set $\left \{ \left < g_{\mu\nu} \right >_k
;\; 0 \leq k< \infty\right \}$ of ordinary metrics\footnote{It 
has been shown \cite{oliverfrac} that in asymptotically safe theories
of gravity, at sub-Planckian distances, spacetime is indeed a fractal 
\cite{mandel,avra,nino} whose spectral
dimension \cite{avra} equals 2.}.
Thus the picture of a ``QEG spacetime" which arises from the 
effective field equations is that of a single 
differentiable manifold equipped with infinitely many
Riemannian structures which are governed by the RG equations. \\

\noindent{\bf 2.3. $k$-independent vs. $k$-dependent diffeomorphisms}

Let us denote the manifold on which the  $\langle g_{\mu\nu}\rangle_k$'s are defined by
$\mathcal{S}$. We interpret $\mathcal{S}$ as a dynamical spacetime and its elements
$P$, $P'$, $\cdots$ as ``events". Let us first focus on the manifold structure of 
$\mathcal{S}$, leaving aside the Riemannian structure for a moment.

On the mathematical side, we introduce local coordinates on $\mathcal{S}$ which establish a 
one-to-one correspondence between points $P$, $P'$, $\cdots$ and coordinate values
$x^\mu(P)$, $x^\mu(P')$, $\cdots$ . On the physical side, we assume that there exists an 
operational procedure which allows us to identify points on $\mathcal{S}$ by means of a
well-defined set of experiments. This procedure for identifying spacetime points is 
required to exist independently of the metric structure of $\mathcal{S}$ so that we can
identify the points of $\mathcal{S}$ in a $k$-independent way. As a consequence, after
having introduced coordinates, we can label the points by $x^\mu(P)$, $x^\mu(P')$, 
$\cdots$, and these labels are the same for all scales.

An example of a simple (thought)
experiment for identifying points could be as follows. Let
$\{ \Phi^a(x)$, $a=1,...,4 \}$ be a configuration of 4 real scalar fields for which the 
map $x^\mu \mapsto \Phi^a$ is invertible. If this configuration is realized on $\mathcal{S}$,
the result $(\Phi^1,\;\Phi^2,
\;\Phi^3,\;\Phi^4)$ of a measurement of all scalars at the same point identifies this point
uniquely\footnote{We are grateful to Max Niedermaier for a helpful discussion of this 
issue.}. 

In the present setting the coordinate system 
(more precisely: the atlas) plays a much more important role than in classical
gravity. It is precisely what can be $k$-independently imposed on spacetime.
One may visualize the whole quantum spacetime as a five-dimensional manifold,
with $k$ labelling the fifth dimension. The 4D coordinates are used to relate
the four-dimensional slices at different $k$-values with each other by saying that
the point on the ($k=k_1$)-slice with coordinates $x^\mu$ is the ``same" as
the point on the ($k=k_2$)-slice with coordinates $x^\mu$. (This is reminiscent of
the 3+1 split in
canonical gravity, where points on different spatial hypersurfaces are 
related via lapse and shift functions.) 

The fact that both the association of coordinates to points and the physical 
identification of points themselves is done in a $k$-independent way has a consequence
which is of crucial importance for the interpretation of the theory: It implies that 
the group of gauge transformations (under which ``physics" is
invariant) consists of {\it $k$-independent} diffeomorphisms $x^\mu \mapsto x'^\mu(x)$
only. As we agreed to use the same coordinate system on $\mathcal{S}$ for all scales
$k$, $k$-dependent coordinate changes $x^\mu \mapsto x'^\mu(x;k)$ are not permitted
since they would alter the relationship between coordinates and physical points
which had been fixed once and for all.

In the family (\ref{fe}) of effective field equations, $k$ plays a purely parametric role,
they do not contain any derivatives with respect to $k$. The equations for different 
$k$-values are decoupled therefore and can be solved for each $k$ separately. Since
$\Gamma_k[g]$ is a diffeomorphism invariant functional, the effective field equations
can determine $\langle g_{\mu\nu}\rangle_k$ at most up to a {\it $k$-dependent}
coordinate transformation. Thus, since the group of gauge transformations consists
of $k$-independent transformations only, it follows that {\it the effective field equations
cannot determine the gauge invariant contents of $\langle g_{\mu\nu}\rangle_k$
uniquely.}

The origin of this non-uniqueness is easy to understand: At one, and only one scale $k$
we have the freedom to fix a gauge, i.e. 
to pick any system of coordinates we like and to express the metric,
at this scale, in terms of those coordinates. But after that the relationship between 
coordinates and physical points (events) is fixed. As a result, from the point of view
of any other scale $k' \neq k$, the coordinates have a physical meaning now and may not
be changed at will any more. The family of field equations (\ref{fe}) cannot ``know"
about this physical interpretation the coordinates have acquired because it is covariant
under a set of transformations which is infinitely much larger than the actual gauge
group, namely under all $k$-dependent diffeomorphisms. This makes them ``blind" to that
part of the difference between $\langle g \rangle_k$ and $\langle g \rangle _{k'}$,
$k' \neq k$, which is due to a change of coordinates. In the later sections of this paper
we shall describe various examples which illustrate this phenomenon.

A different source of ambiguities which the effective field equations cannot resolve are
$k$-dependent constants of integration. If, in classical gravity, a set of solutions
is labeled by one real parameter, say, this parameter will be promoted to a real function
of $k$. Below we shall discuss the example of the Schwarzschild mass $M$ which becomes
a function $M(k)$ in quantum gravity.

If one wants to determine the metrics $\{ \langle g_{\mu\nu}\rangle_k$; $0 \leq k <
\infty \}$ unambiguously one must work with the state $| \Psi\rangle$ or the 
corresponding path integral directly. Given $| \Psi\rangle$ one can, at least in 
principle, derive a path integral whose boundary conditions encode this state,
then add the mode suppression term to it, and follow the standard average action 
construction. Given this path integral, $\langle g_{\mu\nu}\rangle_k$ is unambiguously
defined by $\int \mathcal{D} \gamma \; \gamma_{\mu\nu}(x)\exp(\cdots)$ where all metrics 
refer to the same coordinate system then.\\

\noindent{\bf 2.4. Symmetries and Killing vectors}

Let us assume we are given a family of metrics $\{ \langle g_{\mu\nu} \rangle_k \}$,
all expressed in terms of one and the same system of coordinates on $\mathcal{S}$.
We can now analyze the symmetries of the Riemannian manifold $(\mathcal{S},
\langle g_{\mu\nu} \rangle_k )$ for each value of $k$ separately. We start at some 
$k=k_0$ and search for solutions of the Killing equation
\begin{equation}
 \mathcal{L}_K \langle g_{\mu\nu} \rangle_k =0   
\end{equation}
where $\mathcal{L}_K$ denotes the Lie derivative with respect to the vector field 
$K \equiv K^\mu \partial_\mu$. Let us assume we find a set of Killing vectors
$K_a^\mu$, $a=1,2,\cdots$. They generate the isometry group of $(\mathcal{S},
\langle g_{\mu\nu} \rangle_k )$ at $k=k_0$ in the usual way. If the scale $k_0$
is ``generic" then, by continuity, we expect that $\langle g_{\mu\nu} \rangle_k$
will have the same isometry group also at other scales close to $k_0$. If 
$K_a^\mu \partial_\mu$, for $a$ fixed, is a Killing vector of $\langle g_{\mu\nu} \rangle_k$
on a certain $k$-interval we should distinguish the following situations:\\
\\
(a) The Killing vector is the same for all values of $k$, i.e. $K_a^\mu (x) \partial_\mu$
is independent of $k$.\\
(b) The Killing vector does depend on the scale, i.e. $K_a^\mu (x;k)\partial_\mu$
has an explicit parametric dependence on $k$.\\
\\
In the first (second) case we say that the symmetry is implemented in a 
$k$-independent ($k$-dependent) way. The motivation for this distinction is as follows.
The vector field $K_a^\mu (x;k)\partial_\mu$ generates a flow on $\mathcal{S}$ along
which $\langle g_{\mu\nu} \rangle_k$ does not change. In the case (a) this flow is the 
same for all $k$. In view of the scale independent one-to-one correspondence between
physical points and coordinates this implies that in case (a) the Killing vectors define 
a consistent map of physical points onto physical points.

To be more precise let us consider the two neighboring points on the same ``flow line"
of $K_a^\mu$ which have coordinates $x^\mu$ and $\bar{x}^\mu = x^\mu + \varepsilon
K_a^\mu(x)$, $\varepsilon\ll 1$, respectively. If $K_a^\mu$ is $k$-independent, the map
of coordinates which it induces, $x^\mu \mapsto \bar{x}^\mu$, corresponds to a map
$\mathcal{S}\rightarrow\mathcal{S}$ relating physical points. In case (b) instead, when
$K_a^\mu$ depends on $k$, the target coordinate $\bar{x}^\mu \equiv\bar{x}^\mu(k)$
corresponds to different physical points for different values of $k$. Thus we see that 
{\it if a group of spacetime symmetries is implemented in a $k$-dependent way it no
longer corresponds to a transformation group acting on the manifold of physical events.}
Below we shall discuss concrete examples of both case (a) and (b), respectively.
At certain (non-generic) critical values of $k$ the number of Killing vectors and their
character, in particular the Lie algebra they span, can change discontinuously. \\

\noindent{\bf 2.5. Causal structures}

For any two events $P_1$ and $P_2$ on $\mathcal{S}$ we would like to know whether
$P_1$ can influence $P_2$, or $P_2$ can influence $P_1$, or whether they cannot
influence one another at all. The set of all such relationships between pairs of events 
constitutes a causal structure on $\mathcal{S}$. Within QEG this structure is,
in principle, to be determined as follows. 

Let us consider gravity coupled to some set of matter fields $\chi_I$ and let us fix some
solution $\Gamma_k[g_{\mu\nu},\chi_I]$ of the RG equation. Furthermore, we pick a solution
$\{\langle g \rangle_k, \langle \chi_I \rangle_k$; $0\leq k <\infty \}$
of the resulting coupled effective field equations 
of the gravity plus matter system. This solution 
may be thought of as being implied by some (unperturbed) state
$|\Psi\rangle\equiv |\Psi_{\rm gravity}\rangle |\Psi_{\rm matter}\rangle$. Next one studies
perturbations about this ``vacuum" state by analyzing the properties of the effective
graviton propagator
\begin{equation}
 \left ( \frac{\delta ^2 \Gamma_k}{\delta g_{\mu\nu}(x)\delta g_{\rho\sigma}(y)}
 [\langle g \rangle_k, \langle \chi \rangle_k] \right )^{-1}   
\end{equation}
and the effective matter field propagator
\begin{equation}
 \left ( \frac{\delta ^2 \Gamma_k}{\delta \chi_I(x)\delta \chi_J(y)}
 [\langle g \rangle_k, \langle \chi \rangle_k] \right )^{-1} .     
\end{equation}
Using these effective propagators one then determines the propagation characteristics
of the graviton and matter field fluctuations in the given background. Knowing them,
we can infer which events can be connected to a given event $P_1$ by a propagating gravity
or matter perturbation. For each type of propagating modes one can determine a 
mode-dependent ``causal future" of $P_1$, the set of events which can be influenced by 
$P_1$. Generically these sets will all be different; typically some $P_2$ can be in the 
``causal future" of $P_1$ with respect to one mode but not to some other. 

An example in classical relativity is provided by two events on the light cone of
Minkowski space: if the propagators are the classical ones, the two events can be
connected by the propagating modes of the electromagnetic field, but not of some
massive vector field. In quantum gravity, in particular when the renormalization
effects are strong, the situation is much more involved since the effective propagators
can differ quite substantially from their familiar second-order form because generically
$\Gamma_k$ contains all sets of higher derivative and non-local terms. The graviton propagator
in the fixed point regime \cite{oliver2} is of the $1/p^4$-type, for instance. Moreover,
it is well known that even on a fixed classical spacetime manifold matter quantum effects
alter the propagation characteristics of the matter fluctuations, the photon in particular,
and modify the light cone structure \cite{shore}.

After having defined the mode-dependent ``causal future" of $P_1$ it seems plausible to
define the {\it true causal future} of this point as their union. Every $P_2$ in the 
true causal future of $P_1$ can be influenced by $P_1$ by at least one type of
propagating mode, but typically not all of them.
Analogous remarks apply to the causal past of $P_1$. In the standard situation with
classical second-order propagators all massless fields are equally ``efficient" in 
establishing causal links, they define the boundary of the causal future, while all massive
fields are less efficient. But, as we emphasized above, RG effects can
change this simple pattern.

The upshot of the above discussion is that in QEG the notion of causality is an a
priori scale dependent concept. Its $k$-dependence stems from two different sources:\\
\\
a) The unperturbed metric $\langle g_{\mu\nu}\rangle_k$ is $k$-dependent.\\
b) The propagation characteristics of the field perturbations which are used to 
send signals from one point to another are $k$-dependent.\\

For simplicity's sake, and in order to disentangle the two effects, we shall focus
on the mechanism (a) in the present paper. We assume a situation in which the relevant
propagators are still sufficiently close to the standard second-order ones so that 
the causal structure is determined by the light cones which are implied by the propagators
of the massless fields (the photon propagator in particular, of course) or, in a 
geometric-optical approximation, by the null geodesics. Already in this situation
a remarkable phenomenon arises: the causal structure is scale dependent because the metric
$\langle g_{\mu\nu}\rangle_k$, and therefore the null geodesics it gives rise to, is
$k$-dependent. 

At first a scale-dependent causal structure might appear rather ``exotic" and one might 
wonder whether it can lead to any logical paradoxes. However, its physical origin is easy to 
understand and a tentative interpretation would be as follows. If an event $P_1$ is
the ``cause" of an ``effect" at $P_2$ it must be possible to send a signal from $P_1$ to
$P_2$. Assuming that this signal is transmitted via some field quanta carrying energy
and momentum, the signal itself influences the gravitational field. 
In general it will also be modified by the physical (i.e. gravitating) apparatuses
used as a ``transmitter" and ``receiver".
Within the effective
field theory approach the dominant modifications can be taken into account by changing
the scale at which $\Gamma_k$ is evaluated; the new $k$-value should take into account the 
typical scale set by the signal transmission process. 
Whether or not this is a quantitatively good approximation
depends on how well the transmission of the signal can be modeled by a single-scale
process. Nevertheless, in priciple it is conceivable that different
experimental setups consisting of a transmitter, the signal, and a receiver, involve 
different typical scales (sizes, momenta, virtualities, etc.) and ``see" different 
average metrics $\langle g_{\mu\nu}\rangle_k$ therefore.

It thus can happen that $P_1$ and $P_2$ are on the same side of
an event or particle 
horizon for some $k_1$, while they are on opposite sides for 
some other $k_2$. We
shall find various examples of this phenomenon below.\\ 

\noindent{\bf\large 3. $k$-microscopes}

We use the term ``$k$-microscope" for an idealized experimental set-up, designed to 
observe the structure of the quantum spacetime, whose observations are well described
by the effective field theory provided by the action $\Gamma_k$. The idealization 
involved here is that the ``microscope" is assumed to be characterized by a single scale
only so that it is clear which one of the effective field theories $\{\Gamma_k,
\; 0\leq k < \infty \}$ is to be used for its description. This microscope ``sees" a metric
$\langle g_{\mu\nu} \rangle_k$ solving the effective field equations of the corresponding
action functional $\Gamma_k$.

What is the proper resolution $\ell\equiv\ell(k)$
of such a microscope? Or, equivalenty: What is the coarse graining 
length scale $\ell(k)$ over which
the metric is ``averaged" when observed with the $k$-microscope?
The answer is in general complicated, it depends on the details of the experimental set-up.

Here we will use a simple but universal mathematical model of a microscope
\cite{oliverfrac,minang}, closely 
related to the very construction of the effective average action. 
The input data is the set of 
metrics $\left \{ \left < g_{\mu\nu}\right >_k
\right \}$. The idea is to deduce the relation $\ell = \ell(k)$ from the spectral
properties of the {\it scale dependent} Laplacian ${\bf \Delta} _k \equiv D^2 \left ( 
\left < g_{\mu\nu}\right >_k \right )$ built with the solution of the effective
field equation. For every fixed value of $k$, one solves the
eigenvalue problem of $-{\bf \Delta} _k$ and studies the properties of the
eigenfunctions whose eigenvalue is $k^2$, or nearest to $k^2$ in the case of a discrete
spectrum. We refer to an
eigenmode of $-{\bf \Delta} _k$ whose eigenvalue is (approximately)
the square of the cutoff $k$ as a ``cutoff mode" (COM) 
and denote the set of all COMs by {\sf COM}($k$).

If we ignore the $k$-dependence of ${\bf \Delta} _k$ for a moment (as it would be
appropriate for matter theories in flat space) the COMs are, for a sharp cutoff,
precisely the last modes integrated out when lowering the cutoff, since the 
suppression term in the path integral cuts out all $h_{\mu\nu}$-modes with
eigenvalue smaller than $k^2$.
For a non-gauge theory in flat space the coarse graining (averaging) of fields 
is a well defined procedure, based upon ordinary Fourier analysis.
In this case the length $\ell$
is roughly the wave length of the COMs.

This observation motivates the following 
{\it definition} of $\ell$ in quantum gravity.
We determine the COMs of $-{\bf \Delta} _k$, analyze how fast these eigenfunctions vary
on spacetime, and read off a typical coordinate distance $\Delta x^\mu$
characterizing the scale on which they vary. For an oscillatory COM, for example,
$\Delta x$ would correspond to an oscillation period. 
Finally we use the metric
$\left < g_{\mu\nu} \right >_k$ itself in order to convert $\Delta x^\mu$ to a
proper length. This proper length, by definition, is $\ell$.
Repeating the above steps for all values of $k$, we end up with a function
$\ell =\ell(k)$.
In general one will find that $\ell$ depends on the position on the manifold 
as well as on the direction of $\Delta x^\mu$.

In the following $\ell$ will always denote the intrinsic length scale of the
COMs obtained from the above 
model for a ``$k$-microscope". Our experience with theories in flat spacetime
suggests that the COM scale $\ell$ is a plausible {\it candidate} for a physically 
sensible resolution function $\ell =\ell(k)$, but there might also be others,
depending on the experimental setup one has in mind.

As long as the physical radius of curvature, measured with $\left < g_{\mu\nu} \right >_k$,
is much larger than $1/k$, one can in general use a WKB approximation of
the mode functions to show that $\ell(k)$ is roughly given by 
the classical result $\pi/k$. This $\ell(k)$ is a 
{\it proper} length measured with $\left < g_{\mu\nu} \right >_k$.
The {\it coordinate} distance $\Delta x$ 
from which $\ell$ was obtained (and the 
proper length obtained when this $\Delta x$ is ``measured" with a fixed macroscopic
metric $\left < g_{\mu\nu} \right >_{k_{\rm lab}}$) may depend on $k$ in a completely 
different way. In ref.~\cite{minang} we showed that, in the case of a Euclidean
four-sphere, there is a minimal coordinate distance $\Delta x$ with the property that
a $k$ exists so that $\Delta x$ can be resolved by the corresponding cutoff modes.
This is true although there is no lower bound on $\ell(k)$, which runs $\propto 1/k$
all the way towards $k \rightarrow\infty$.

In the presence of strong curvature, $\ell(k)$ may deviate substantially from $\pi/k$.
It is then in general necessary to write $\ell(k,x,n)$ to account for the
dependence on position and direction. We specify the direction by a unit vector $n^\mu$. 

The $n^\mu$-dependence is particularly important
if the signature of the metric is Lorentzian. Because of the 
possible compensation of timelike and spacelike oscillations, 
one could then have arbitrarily small $\ell$ for arbitrarily small $k$.
In flat Minkowski space, say, a wave $\sim e^{i(\vec{k}\vec{x}-|\vec{k}|t)}$ has
$k^2=0$ but the wavelength $2 \pi/|\vec{k}|$ can be arbitrarily small.
A working procedure to determine $\ell(k,x,n)$ 
for a given direction $n^\mu$ would be to define it from the COMs
which vary as little as possible in any direction orthogonal to $n$. \\

\noindent{\bf\large 4. The Schwarzschild-de~Sitter family}

In this section we illustrate several of the points discussed above by means of 
explicit examples. We consider pure 
Lorentzian gravity in the Einstein-Hilbert approximation.
The family of effective field equations is given by eq.~(\ref{fieldeq}) then.\\

\noindent{\bf 4.1. Running metric for a generic state}

Let us find the most general solution to eqs.~(\ref{fieldeq}) with $\Lambda(k)>0$
subject to the symmetry constraint that  $\langle g_{\mu\nu}\rangle_k$ is
spherically symmetric (isotropic) and stationary on all scales. On the constant-time
surfaces we use polar coordinates $r,\theta,\phi$, and we define the time coordinate 
such that $\partial/\partial t$ is the Killing vector related to stationarity.

Applying the familiar textbook arguments \cite{weinbook} at each
value of $k$ we see that in these coordinates the most general static isotropic
line element $\langle ds^2 \rangle_k \equiv  \langle g_{\mu\nu}\rangle_k dx^\mu dx^\nu$
is given by 
\begin{eqnarray}\nonumber
 \langle ds^2 \rangle_k = &-F(r;k)dt^2 + 2r E(r;k)dt dr + r^2 D(r;k)dr^2 \\ 
 \label{s1}
 & +C(r;k) [dr^2+ r^2 d \theta^2 + r^2 \sin^2 \theta d \phi^2].  
\end{eqnarray}
It contains 4 free functions, $C$, $D$, $E$ and $F$, which depend on the coordinate
$r$ and the parameter $k$.

Let us first recall the situation in classical gravity where the 4 functions are
$k$-independent. There one can perform a change of coordinates which reduces the 
number of free functions to 2. If one introduces as new coordinates
\begin{equation}\label{s2}
 r'= r \, C(r)^{1/2}, \quad\quad t'=t+T(r)   
\end{equation}
with $T(r)=-\int dr \; r E(r)/F(r)$, then the line element can be brought to the
``standard form"
\begin{equation}\label{s3}
 ds^2=-B(r')\, dt'^2+ A(r')\, dr'^2+ r'^2 \, [d \theta^2 +\sin^2 \theta d \phi^2].   
\end{equation}
The two new coefficient functions $A$ and $B$ can be expressed in terms of the original 
ones, see ref.~\cite{weinbook}. 

In quantum gravity where the metric coefficients depend on $k$ an analogous reduction
from 4 to 2 free functions is not possible. The reason is that in this case the 
change of coordinates (\ref{s2}) would involve $k$-dependent functions $C(r;k)$ and
$T(r;k)$ but, as we discussed in section 2.3, the group 
of gauge transformations consists of $k$-independent
diffeomorphisms only. We would like the coordinates $(t_1,r_1,\theta_1,\phi_1)$ to
belong to the same physical point $P_1$ for all values of $k$, and therefore only
$k$-independent coordinate transformations are possible. We may use the freedom to perform
$k$-independent transformations in order to transform $\langle ds^2 \rangle_k$ to 
the standard form (\ref{s3}) at one particular value of $k$ at most. The gauge transformations
are ``used up" then, and on all other scales $k' \neq k$ one still needs 4 functions
$C$, $D$, $E$ and $F$ to parametrize the most general static isotropic metric.

The parametrization (\ref{s1}) is now used as an ansatz for solving the familiy of
field equations. Inserting (\ref{s1}) into (\ref{fieldeq}) one obtains a coupled 
system of ordinary differential equations, involving $C$, $D$, $E$ and $F$, and its
derivatives with respect to $r$. We shall not write down these 
complicated equations here. 
Suffice it to say that the equations belonging to different $k$-values are completely
decoupled and can be solved for each $k$ separately. But at fixed $k$ the situation is the 
same as in classical gravity where the field equations determine only two functions,
after the other two have been fixed by an appropriate choice of coordinates.

Here we see very explicitly that the field equations cannot fix $\langle g_{\mu\nu}\rangle_k$
completely. As we discussed already, they determine $\langle g_{\mu\nu}\rangle_k$ up to
a $k$-dependent diffeomorphism. But since the group 
of gauge transformations consists of $k$-independent 
diffeomorphisms only, the family $\{ \langle g_{\mu\nu}\rangle_k$; $0\leq k < \infty \}$
encodes additional physical information which is ``known" to the state $|\Psi\rangle$ 
only, but not to the effective field equations. \\

\noindent{\bf 4.2. A special class of states}

The previous subsection can be summarized by saying that in the static isotropic case
the running metric is parametrized by 4 functions of $r$ and $k$. The effective
field equations allow us to express 2 of them in terms of the other 2; the latter can be found 
only from $|\Psi\rangle$ or the corresponding path integral directly.

In order to illustrate another point we shall now assume that there exists a class of 
states $|\Psi\rangle$ for which, for any $k$, the metric assumes the more special form
\begin{equation}\label{s4}
 \langle ds^2 \rangle_k = -f(r;k) dt^2 + \frac{dr^2}{g(r;k)}+ r^2 d\theta^2
 + r^2 \sin^2 \theta d\phi^2.   
\end{equation} 
We emphasize that (\ref{s2}) represents an {\it assumption}, there is no guarantee that
a state with this property exists. Taking the more restricted structure (\ref{s4})
for granted, the effective field equations determine the functions $f$ and $g$ 
almost completely. Inserting (\ref{s4}) into (\ref{fieldeq}) one infers that the most
general solution has $g(r;k)=f(r;k)$ and
\begin{equation}\label{s5}
 f(r;k)=1-\frac{2 m(k)}{r}-\frac{\Lambda(k)}{3} r^2   .
\end{equation}
Obviously, for every $k$, (\ref{s4}) with (\ref{s5}) is a Schwarzschild-de~Sitter
metric for the cosmological constant $\Lambda(k)$. Here $m(k)$ is a constant of
integration which may be chosen differently at different scales. While $\Lambda(k)$
is dictated by the RG equation, we cannot deduce $m(k)$ from the effective field equations.

This is an example of the second source of ambiguities mentioned in subsection 2.3: 
Constants of integration which are just numbers in classical gravity become functions
of $k$ in quantum gravity. This type of ambiguity, too, can be resolved only by
analyzing the state $|\Psi\rangle$ directly.

As in classical gravity one might interpret $m(k)\equiv G(k)M(k)$ as the product 
of the running Newton constant and a running black hole mass $M(k)$. From the point
of view of the effective field equations this seems a bit artificial, though, since
in vacuo they do not contain $G(k)$. 

The function $f(r;k)$ is positive only in a finite portion of spacetime. This region is 
sandwiched between two horizons at which $f(r;k)=0$: 
the black hole event horizon at $r=r_{\rm e}(k)$ and the
cosmological horizon at $r=r_{\rm c}(k)$. As long as $m(k)$ and $\Lambda(k)$ are both 
sufficiently 
small, we have $r_{\rm e}(k) \ll r_{\rm c}(k)$, and their values are approximately
\begin{equation}
 r_{\rm e}(k)\approx 2 m(k), \quad r_{\rm c}(k)\approx\sqrt{\frac{3}{\Lambda(k)}}.
\end{equation}

In order to get a first understanding of the scenarios which are in principle possible
for the ``zooming" into the QEG spacetime we consider two special examples for the 
function $m(k)$. We assume that there exist states $|\Psi\rangle$ giving rise to this 
particular form of $m(k)$, but strictly speaking there is no guarantee for their existence.
However, the first example is well motivated both by explicit perturbative and RG
computations which establish the postulated behavior in a certain regime at least.\\
\newpage
\noindent {\bf (a) $m(k)$ decreases with growing $k$}.

We assume that, in a certain interval of $k$-values, (i) the function $m(k)$ decreases
monotonically with increasing $k$, and (ii) the cosmological constant $\Lambda(k)$ is small
there so that $r_{\rm c}(k)\gg r_{\rm e}(k)$. 
The second assumption means that we are essentially
looking at a Schwarzschild black hole for which the cosmological horizon plays no 
important role.

Writing $m(k)\equiv M G(k)$ with $M=const$ we see that the assumption (i) amounts to 
a running $G(k)$ which decreases with increasing $k$. This is exactly the ``antiscreening"
behavior implied by the RG equations \cite{mr}. In fact, in refs.~\cite{bh} the 
RG-improved Schwarzschild metric was obtained by taking the running of Newton's 
constant into account, and contact was made with the quantum corrected Newtonian potential
calculated perturbatively within the effective field theory approach to gravity
\cite{donoghue,bohr}. These explicit analyses indicate that there should indeed exist
states with the properties assumed above.

Since $r_{\rm e}(k)\approx 2m(k)$, a decreasing $m(k)$ implies that the event horizon moves
{\it inward} as we {\it increase} the scale $k$. As a result, there exist (space) points
$Q$ which are inside the horizon according to the metric $\langle g_{\mu\nu}\rangle_{k_1}$,
but are outside when one uses $\langle g_{\mu\nu}\rangle_{k_2}$ with $k_2>k_1$ instead.
Let us position some observer at a point $Q'$ with $r_{Q'}> r_{\rm e}(k_1)$. 
Then this observer 
can receive signals from $Q$ according to the causal structure pertaining to the scale $k_2$,
but not to the one for $k_1$. This is an example of ``scale-dependent causality".

While seemingly paradoxical at first sight, its interpretation should be clear from 
the discussion above: In the quantum regime, the transmission of a signal can no longer
be modeled by a (massless) test particle, but rather the backreaction 
onto the metric of the complete
physical system consisting of a transmitter, the signal per se, and a receiver 
has to be taken into account. In the average action approach this is done,
to leading order, by changing the relevant scale $k$.

In describing the transmission process by the set of metrics (\ref{s4}) we made the tacit
assumption that it preserves the original symmetries of the spacetime without the
transmission apparatus. This might be an oversimplification when it comes to describing 
realistic physical experiments, but it does not affect our general conclusion that 
causality is a resolution dependent notion, in principle. \\

\noindent {\bf (b) $m(k)$ is constant.}

Next we assume a state for which the function $m(k)\equiv m_0$ 
is constant and take $G(k)$ and $\Lambda(k)$ to follow a Type IIIa trajectory.
Both horizons are now unaffected for $k \apprle k_T$, 
where $k_T$ is the turning point of the
trajectory, as described in the Introduction. For $k$ well above $k_T$, $\Lambda(k)$  
grows $\propto k^4$. This implies that the radius of the cosmological horizon gets smaller
as $k$ increases: $r_{\rm c}(k)\sim k^{-2}$. 
(This is analogous to the $k^{-2}$-shrinking of the sphere in \cite{minang}.)
At the same time, the inner horizon at $r_{\rm e}(k)$ is driven 
outwards because of the growing importance
of the term $\frac{\Lambda(k)}{3}r^2$ in $f(r;k)$. Finally, at $k=k_{\rm m}$, the two horizons merge
when $\Lambda(k)$ reaches the value
\begin{equation}
 \Lambda(k=k_{\rm m})=\frac{1}{9 m_0^2}, 
\end{equation} 
at the position
\begin{equation}
 r_{\rm e}(k_{\rm m})=r_{\rm c}(k_{\rm m})=3 m_0=\frac{1}{\sqrt{\Lambda(k_{\rm m})}}.   
\end{equation}
The scale $k_{\rm m}$ where the horizons 
merge can be in the range where $\Lambda(k)\sim k^4$ or in the fixed point
regime where $\Lambda(k)\sim k^2$, depending on 
the value of $m_0$. For $k>k_{\rm m}$, $f(r)$ is negative
everywhere, and as a result the causal structure of spacetime on these 
scales is completely different from that at small $k$. \\

\noindent{\bf\large 5. Robertson-Walker families}

Next we analyze the effective field equations imposing a different symmetry requirement.
We assume that, for any value of $k$, the effective spacetime is foliated by spacelike
hypersurfaces which are homogeneous and isotropic. Specializing further for flat
hypersurfaces, the most general metric consistent with these requirements is of the form
\begin{equation}\label{cosmet1}
 \langle ds^2 \rangle_k \equiv \langle g_{\mu\nu} \rangle_k \, dx^\mu dx^\nu 
 =-b^2(t;k)\, dt^2 + a^2(t;k)\,\delta_{ij}dx^i dx^j.   
\end{equation}
It contains two free functions $a$ and $b$ of the time coordinate $t$. If it were not
for their parametric $k$-dependence, one could redefine the time variable
$(t \rightarrow t'=\int b\,dt)$ in order to achieve $b=1$. This would lead us to the standard
form of the Robertson-Walker line element as it is usually used in cosmology. However, 
as the group of gauge
transformations consists of $k$-independent diffeomorphisms only, $b=1$
can be achieved for one value of $k$ at most, but not for all. Therefore, if the
symmetry requirement is the only subsidiary condition constraining the form of
$\langle g_{\mu\nu} \rangle_k$, the most general form of the running metric contains two 
free functions of $t$ and $k$.

For generic functions $a$ and $b$ the metric (\ref{cosmet1}) admits 6 Killing vectors
$K_a^\mu \partial_\mu$ related to homogeneity and isotropy. They are {\it $k$-independent}
since they just translate and rotate the spatial 
cartesian coordinates $x^i$ in the usual way; these 
transformations do not involve the $k$-dependent functions $a$ and $b$ as the $x^i$'s
do not get mixed with $t$.\\

\noindent{\bf 5.1. Vacuum solutions}

Again employing the Einstein-Hilbert truncation, we assume that no matter is present
so that the effective Einstein equations assume the form (\ref{fieldeq}). Inserting
the ansatz (\ref{cosmet1}) we obtain a single differential equation for $a$ and $b$
\footnote{Here and in the following a dot denotes a derivative with respect to $t$.}:
\begin{equation}\label{f1}
 3 \left ( \frac{\dot{a}(t;k)}{a(t;k)}\right )^2 = \Lambda(k)\, b^2(t;k).  
\end{equation}
Here we encounter another example of the phenomenon that the effective field equations
cannot fully determine the physical contents of the running metrics because they are
invariant under $k$-dependent diffeomorphisms while physics is invariant under 
$k$-independent ones only. The single equation (\ref{f1}) does not contain enough
information to determine both $a(t;k)$ and $b(t;k)$ after fixing initial conditions. 
The reason is clear: By a $k$-dependent change of the time coordinate we can transform
$b$ into any function we like, including $b \equiv 1$, if we use the new time
\begin{equation}\label{f2}
 t'(t;k)=\int^t b(\bar{t};k) d \bar{t}.   
\end{equation} 
But this transformation is forbidden in the present context. The new time coordinate
depends on the old one in a scale dependent way which destroys the required
$k$-independent one-to-one correspondence between coordinates and physical points.

The allowed transformations are, however, sufficient to achieve $b=1$ at a single
scale, $k=k_0$ say. Fixing the ($k$-independent!) coordinate system in this way we can
then use (\ref{f1}) to determine $a$ at $k_0$ uniquely (up to initial conditions).
The result reads
\begin{equation}\label{solcos1}
 a(t;k_0)=\exp \left [ \pm \sqrt{\frac{\Lambda(k_0)}{3}}(t-t_1)\right ] ,\quad b(t;k_0)=1   
\end{equation}
where $t_1$ is an integration constant. Without further information which must come from
the state directly it is impossible to determine $a(t;k)$ and $b(t;k)$ for $k \neq k_0$.

The metric (\ref{cosmet1}) at $k_0$, with the functions (\ref{solcos1}) describes a 
patch of de~Sitter space. As a result, $\langle g_{\mu\nu}\rangle_{k_0}$ has more symmetries
than those built into the ansatz. It admits 10 rather than just 6 Killing vectors.

At $k\neq k_0$ the state $|\Psi\rangle$ dictates a certain function $b(t;k)\neq 1$. 
Nevertheless, the metric $\langle g_{\mu\nu}\rangle_k$ still admits 10 Killing vectors.
The reason is that the condition $\mathcal{L}_K \langle g_{\mu\nu}\rangle_k =0$ is
covariant under $k$-dependent diffeomorphisms. As a consequence, the (non-)existence
of Killing vectors cannot depend on whether we use $t$ or $t'(k)$ as a time coordinate.
Hence $\langle g_{\mu\nu}\rangle_k$ given by (\ref{cosmet1}) with functions $a$ and $b$
constrained by the differential equation (\ref{f1}) but arbitrary otherwise is maximally
symmetric for {\it any} value of $k$. 

The 6 Killing vector fields related to homogeneity and isotropy and the 4 additional
ones responsible for the enlarged symmetry are not on an equal footing, though. While
the former are $k$-independent, the latter may depend on the scale. The coordinate
transformations generated by the former do not mix the $x^i$'s with $t$, but the latter
do. As a result, the components $K_a^\mu(x;k)$ of the 4 additional 
Killing vectors depend on $k$
explicitly since they ``feel" the $k$-dependent coefficient functions $a$ and $b$ in
(\ref{cosmet1}). According to the discussion in section 2.4 this implies that the 
original 6 Killing vectors can be regarded transformations on the manifold $\mathcal{S}$
of physical events, but not necessarily the 4 new ones. We observe a kind of symmetry 
breaking here; it represents an ``anomaly" in the sense that it is caused by quantum
effects. 

This anomaly occurs already if there are states with $b(t;k)=1$ for all $t$ and $k$.
The familiar metric
\begin{equation}\label{cosmet2}
 \left < ds^2 \right >_k = -dt^2+a(t;k)^2 dx^i dx^i   
\end{equation} 
with the scale factor
\begin{equation}\label{adesit}
 a(t;k)=A \exp\left [ \sqrt{\frac{\Lambda(k)}{3}}\, t \right ]   
\end{equation}
is well known to be maximally symmetric, but even with a $k$-independent constant
of integration, $A$, some of the Killing vectors are unavoidably scale dependent
because the cosmological constant is. 

The cosmology given by (\ref{adesit}) has a
running Hubble parameter $H(k)=\sqrt{\Lambda(k)/3}$. Since $\Lambda$ is a monotonically
increasing function of $k$, so is $H(k)$. As a result, a high-resolution microscope
will see the universe expand faster than one with a poorer resolving power. In the 
fixed point regime where $\Lambda(k)\propto \lambda_* k^2$, say, we have
\begin{equation}\label{adesitfp}
 a(t;k)=A \exp \left [ \sqrt{\frac{\lambda _*}{3}}\, kt \right ]  
\end{equation}
with $H(k)$ directly proportional to $k$.
\\

It is natural to search for a state $|\Psi\rangle$ in which the full de~Sitter symmetries
are $k$-independently realized. Only such a state, with 10 $k$-independent Killing vector
fields, would we call a {\it Quantum de~Sitter space}, since only then spacetime
is maximally symmetric as a manifold $\mathcal{S}$ of physical
events. What are the conditions $a(k;t)$
and $b(t;k)$ must obey in this case?

The $k$-dependence of the functions $a$ and $b$ must reflect the maximal symmetry 
of the quantum space time. They have to grow or shrink in the same way as functions of $k$,
and this growing or shrinking has to be independent of the position. Taking again $k_0$ as
a reference scale, these requirements imply
\begin{equation}\label{j1}
 \frac{a(t;k)}{a(t,k_0)}=\frac{b(t;k)}{b(t;k_0)}=c(k),   
\end{equation}
where $c(k)$ is a function of $k$ only.

To see the necessity of eq.~(\ref{j1}) for the full symmetries to hold $k$-independently,
we note that maximally symmetric (classical) spacetimes have the following property: 
Let $u$ be a vector in the tangent space of a point $x$ and $v$
a vector in the tangent space of a point $y$. If $u$ and $v$ have the same length,
then there is {\it always} an isometry transformation which maps $x \mapsto y$ and
$u \mapsto v$ (up to a time reversal). 
In the quantum case, if all isometries are required $k$-independent,
we therefore must demand that if $u$ and $v$ have the same length at {\it some} 
scale $k_0$, they automatically have the same length at {\it all} scales $k$, since
they are always linked by the same isometry transformation. This implies that they obey
\begin{equation}
 \langle g_{\mu\nu}(x)\rangle_k \, u^\mu u^\nu = 
 \langle g_{\mu\nu}(y)\rangle_k \, v^\mu v^\nu 
\end{equation}
for all $k$. Since $u$ and $v$ generically point into different
directions (e.g. if they are related by an isometry which is 
a combination of translations, rotations and boosts) this is only possible when all
components of the metric have the same dependence on $k$, and when this $k$-dependence
is independent of the position. Otherwise $u$ and $v$ would shrink or grow differently
when $k$ is changed, and they would be no longer of equal length. 
This proves eq.~(\ref{j1}).

Using (\ref{j1}) in the field equation (\ref{f1}), we see that the only solution
for $c(k)$ is
\begin{equation}\label{j2}
 c(k)=\sqrt{\frac{\Lambda(k_0)}{\Lambda(k)}}.    
\end{equation}
This $k$-dependence of the metric is completely identical to what we found for
the four-sphere in ref.~\cite{minang}.      
In combination with eq.~(\ref{solcos1}) we obtain the solution for 
the metrics of quantum de~Sitter space:
\begin{equation}\label{solcos2}
  a(t;k)=\sqrt{\frac{\Lambda(k_0)}{\Lambda(k)}}\exp \left [ \pm \sqrt{\frac{\Lambda(k_0)}{3}}
  (t-t_1)\right ] ,
  \quad b(t;k)=\sqrt{\frac{\Lambda(k_0)}{\Lambda(k)}}.    
\end{equation}
We shall return to this result in section 6.2.\\
\newpage
\noindent{\bf 5.2. Cosmological solutions with matter}

Next we add a matter piece $\Gamma_k^{\rm M}$ to the Einstein-Hilbert ansatz for the
average action. The effective field equations assume the form (\ref{m1}) then
where the scale dependent energy momentum tensor $\langle T^{\mu\nu}\rangle_k$
is given by (\ref{m2}). We impose the symmetry constraint of homogeneity and isotropy again
and consider the case of flat $t=const$ slices. In the adapted $(t,x^i)$-system of 
coordinates, eq.~(\ref{cosmet1}) is the most general metric then and, for symmetry reasons,
$\langle T_\mu \,^\nu \rangle_k \equiv \langle g_{\mu\rho}\rangle_k
\langle T^{\rho\nu}\rangle_k$ has the structure
\begin{equation}\label{f20}
 \langle T_\mu \,^\nu \rangle_k= {\rm diag} [-\rho,\, p,\, p,\, p]   
\end{equation}
with functions $\rho\equiv\rho(t;k)$ and $p \equiv p(t;k)$. If we insert (\ref{f20}) 
and the metric ansatz (\ref{cosmet1}) into the effective Einstein equations (\ref{m1})
we obtain two independent differential equations. We take them to be the modified
Friedmann equation
\begin{equation}\label{friedmann}
 3 \left ( \frac{\dot{a}(t;k)}{a(t;k)}\right )^2 = 
 \bigg[8 \pi G(k)\,\rho(t;k)+\Lambda(k)\bigg]
 \, b^2(t;k)   
\end{equation}
and the continuity equation
\begin{equation}\label{f21}
 \dot{\rho}(t;k)+3 \,\frac{\dot{a}(t;k)}{a(t;k)}\,
 [\rho(t;k)+p(t;k)]=0   
\end{equation}
which corresponds to (\ref{m2a}). These equations are to be supplemented by the matter
equation of motion or, in a hydrodynamical description of the matter system, by an equation
of state $p=p(\rho;k)$.\\

\noindent{\bf 5.2.1. Solutions with $k$-independent causality by constant Weyl rescaling}

In absence of matter there exists a simple general method for generating solutions
$\langle g_{\mu\nu}\rangle_k$ if an initial solution $\langle g_{\mu\nu}\rangle_{k_0}$
at some reference scale $k_0$ is known: it is enough to multiply the initial metric by an
$x$-independent conformal factor \cite{oliverfrac}. In fact,
\begin{equation}\label{f30}
 \langle g_{\mu\nu}(x)\rangle_k=\frac{\Lambda(k_0)}{\Lambda(k)}
 \,\langle g_{\mu\nu}(x)\rangle_{k_0}   
\end{equation}
is a solution to (\ref{fieldeq}) for any $k$ if it is at $k_0$ (excluding, as always,
topology/symmetry changes in the $k$-interval considered). An example is the solution
(\ref{solcos2}). Clearly the family (\ref{f30}) has the same light cone structure at any
$k$ and thus provides an example of scale independent causality.

As we are going to argue that scale independent causality is more the exception than
the rule it is important to understand that in the presence of matter constant Weyl
transforms can be a solution only under highly non-generic and ``unnatural" conditions.

For the metric ansatz (\ref{cosmet1}), eq.~(\ref{f30}) is equivalent to
\begin{equation}\label{f31}
 a(t;k)=\sqrt{\frac{\Lambda(k_0)}{\Lambda(k)}}a(t;k_0), \quad
 b(t;k)=\sqrt{\frac{\Lambda(k_0)}{\Lambda(k)}}b(t;k_0).   
\end{equation}
It is easy to verify that (\ref{f31}) is a solution to (\ref{friedmann}) and (\ref{f21})
provided the energy density and the pressure scale as follows:
\begin{eqnarray}\label{f32}
 \rho(t;k)&=&\frac{\Lambda(k)}{G(k)}\left ( \frac{\Lambda(k_0)}{G(k_0)}\right )^{-1}
 \rho(t;k_0),\\ \nonumber
 p(t;k)&=&\frac{\Lambda(k)}{G(k)}\left ( \frac{\Lambda(k_0)}{G(k_0)}\right )^{-1}
 p(t;k_0)   
\end{eqnarray}
If (\ref{f32}) happens to be satisfied we indeed have found a solution with 
$k$-independent causality. However, the $k$-dependence of $\rho$ and $p$ is dictated
independently by the RG flow of $\Gamma_k^{\rm M}$ so that in general there is no
reason for (\ref{f32}) to hold. Note in particular that (\ref{f32}) implies a very
special ``equation of state"
\begin{equation}\label{f33a}
 p(t;k)=w(t)\rho(t;k)   
\end{equation}
where
\begin{equation}\label{f33b}
 w(t)\equiv p(t;k)/\rho(t;k)   
\end{equation}
depends on time but not on $k$. The factors on the RHS of eqs.~(\ref{f32}) have an
interesting interpretation. Since
\begin{equation}\label{f34a}
 \rho_\Lambda(k) \equiv \frac{\Lambda(k)}{8 \pi G(k)}   
\end{equation}
is the scale dependent (but $t$-independent!) vacuum energy density due to the cosmological
constant we may write $\rho$ and $p$ in a form where its $t$- and $k$-dependencies
factorize:
\begin{equation}\label{f34b}
 \rho(t;k)=\frac{\rho_\Lambda(k)}{\rho_\Lambda(k_0)}\,\rho(t;k_0)   
\end{equation} 
and likewise for $p$. The above solution with $k$-independent causality exists if,
and only if, the matter energy density, at any time, scales with $k$ in the same way
as the vacuum energy density.

These remarks indicate that the relations (\ref{f32}) are highly constraining. We 
illustrate this point for a single scalar field $\chi$ and the ``local potential
approximation" 
\begin{equation}
 \Gamma_k^{\rm M}[g,\chi]=-\int d^4 x \sqrt{-g}\left ( \frac{1}{2}(D_\mu \chi)^2
 +V_k (\chi) \right ).   
\end{equation}
Then, for a spatially homogeneous solution $\langle \chi(t)\rangle_k$,
\begin{eqnarray}\label{f34}
 \rho(t;k)&=&\frac{1}{2 b^2}\left ( \frac{d}{dt}\langle \chi(t)\rangle_k \right )^2
 +V_k(\langle \chi(t)\rangle_k), \\ \nonumber   
 p(t;k)&=&\frac{1}{2 b^2}\left ( \frac{d}{dt}\langle \chi(t)\rangle_k \right )^2
 -V_k(\langle \chi(t)\rangle_k)   
\end{eqnarray}
Here $\rho$ and $p$ have both an explicit $k$-dependence via the RG running of the 
effective average potential $V_k$ and an implicit one via the solution
$(\langle g \rangle_k, \langle \chi \rangle_k)$. It is clear that for a generic RG
trajectory $\{G(k),\Lambda(k),V_k(\cdot)\}$ and generic solution to the 
resulting $t$-dynamics
eqs.~(\ref{f34}) will not comply with (\ref{f32}).

One might try to search for solutions with $k$-independent causality by allowing
the metrics at different scales to be related by a position-dependent conformal
factor:
\begin{equation}\label{f35}
 \langle g_{\mu\nu}\rangle_k= C(x;k,k_0)\langle g_{\mu\nu}\rangle_{k_0}.   
\end{equation}
However, for a generic RG trajectory and solution to the field equations the metric 
will {\it not} be of the form (\ref{f35}). The reason is clear: in the exact theory
the RG trajectory amounts to infinitely many running couplings such as $G(k)$, 
$\Lambda(k)$, or the function $V_k(\cdot)$ which by itself contains already infinitely
many couplings. All of these couplings get changed when we switch to another 
trajectory. At the level of solutions, this change cannot in general be absorbed by a 
change of the single function $C(x;k,k_0)$. Therefore, unless one is dealing with a
highly symmetric theory or performs an extreme finetuning of initial conditions,
a generic solution for the running metric will not be of the type (\ref{f35}). As
a result, the causal structure will depend on $k$ then.\\

\noindent{\bf 5.2.2. Cosmologies with $k$-dependent causality}

In order to illustrate the case of scale dependent causal structures we {\it assume}
in this section that there exists a class of special states with
\begin{equation}\label{f36}
 b(t;k)=1 \quad {\rm for}\; {\rm all}\; t \;{\rm and}\; k.   
\end{equation}
We employ a hydrodynamical description of the matter system and take the equation of state
to be $p=\rho /3$, corresponding to a traceless energy momentum tensor. Hence the metric
reads $\langle ds^2 \rangle_k=-dt^2 + a^2(t;k) d {\bf x}^2$, and the Friedmann equation
(\ref{friedmann}) and continuity equation (\ref{f21}) assume their standard form. For 
an RG trajectory with $\Lambda(k)> 0$ in the $k$-interval of interest their general
solution reads:
\begin{equation}\label{f37}
 a(t;k)=\left [ \frac{\mathcal{M}(k)G(k)}{2 \Lambda(k)}\left \{ \cosh \left [
 \frac{4}{3} \sqrt{3 \Lambda(k)}(t-t_0)\right ] -1 \right \}\right ]^{1/4},    
\end{equation}
\begin{equation}\label{f38}
 \rho(t;k)=\frac{\mathcal{M}(k)}{8 \pi a^4(t;k)}.   
\end{equation}
This solution contains two constants of integration, $\mathcal{M}(k)$ and $t_0$.
The $k$-dependence of $\mathcal{M}(k)$ is not fixed by the Einstein equations.
On the other hand, 
the constant of integration $t_0$ cannot depend on $k$, since this would be inconsistent
with our definition of an effective QEG spacetime: 
The range of the time coordinate is the
interval $(t_0,\infty)$. The universe starts with a big bang singularity at $t=t_0$. 
If we had $t_0(k_1)<t_0(k_2)$, the era between $t_0(k_1)$ and $t_0(k_2)$ would
exist when spacetime is probed at the scale $k=k_1$ but not at $k=k_2$. This would be
in contradiction with our assumption that there is the same 
one-to-one correspondence between 
coordinates and events for all $k$. Thus $t_0$ must be independent of $k$, and we
may set $t_0=0$ by readjusting the time axis.  

Let us consider the fixed point regime as an example. Every QEG trajectory\footnote{
The quantum fluctuations of the matter fields influence the RG flow. The NGFP is known
to persist for a broad class of matter systems, however \cite{perper1}.}
starts near the NGFP $(g_*,\lambda_*)$ where the dimensionful gravitational parameters
behave as
\begin{equation}\label{f39}
 G(k)=g_*/k^2, \quad \Lambda(k)=\lambda_* k^2.   
\end{equation}
With this RG running, valid for $k \apprge m_{\rm pl}$, eq.~(\ref{f37}) becomes
\begin{equation}\label{f40}
 a(t;k)=\frac{1}{k}\left [ \frac{g_* \mathcal{M}(k)}{2 \lambda_*}\right ]^{1/4}
 \left \{ \cosh \left [ \frac{4}{3}\sqrt{3 \lambda_*}\, k \, t \right ] -1 \right \}^{1/4}. 
\end{equation}
It is instructive to analyze (\ref{f40}) in the regimes where $k$ is much smaller
or larger than the second mass scale in the problem, $1/t$.
We have
\begin{equation}\label{f41}
 a(t;k)=\left ( \frac{g_*}{6} \right )^{1/4} \mathcal{M}(k)^{1/4}\sqrt{t/k} 
 \quad {\rm for} \quad m_{\rm Pl}\apprle k \ll 1/t,   
\end{equation} 
\begin{equation}\label{f42}
 a(t;k)= \left ( \frac{g_* \mathcal{M}}{2 \lambda_*}\right )^{1/4}
 \frac{1}{k} \exp(\sqrt{\lambda_*/3}\, kt)
 \quad {\rm for} \quad k \gg 1/t.
\end{equation}
A ``microscope" with a comparatively poor resolution, corresponding to a small $k \ll 1/t$,
sees essentially the classical $a \propto \sqrt{t}$ expansion. Since $1/t \propto H(t)$
here, this microscope focuses on ``super-Hubble" structures. On the other extreme, a high
resolution microscope with $k \gg 1/t$ perceives the universe as exponentially inflating.
Its Hubble parameter $H=\sqrt{\lambda_*/3}\, k$ is constant in time but depends on $k$:
the better the resolution of the microscope is, the faster seems the universe to expand.
This phenomenon is related to the fractal and self-similar properties of QEG spacetimes
discussed earlier \cite{oliver1,oliver2,oliverfrac}. 

At this point we can make contact with the RG improvement approach.
A microscope whose resolution
is continuously readjusted so that $k=1/t$ sees the universe expanding according to
\begin{equation}\label{f43}
 a(t;k)\propto (\mathcal{M}(1/t))^{1/4}\, t \quad {\rm for} \quad 
 k=1/t.  
\end{equation}
For $\mathcal{M}=const$ one recovers precisely the linear expansion $a \propto t$
which was found in \cite{cosmo1} by a completely different reasoning.

The causal structure of the spacetime with the scale factors (\ref{f40}) does indeed
depend on $k$. This becomes manifest when one investigates possible particle horizons,
for example. A Robertson-Walker metric implies a particle horizon of coordinate
(or comoving) radius 
\begin{equation}\label{f44}
 r_H(t;k)=\int_0^t \frac{dt'}{a(t';k)}   
\end{equation}
provided the integral on the RHS of (\ref{f44}) converges at its lower limit.
If $r_H$ is finite, a fixed event at $r=0$ and time $t$ can be influenced causally only
by the events inside a spatial ball with this coordinate radius. Now, since the 
relationship between coordinates and events is strictly $k$-independent, a scale
dependence of $r_H(r;k)$ means that the set of events which can influence the event
at $r=0$ and time $t$ is $k$-dependent. For the $a(t;k)$ of eq.~(\ref{f40}) this is
indeed seen to happen. 

The radius (\ref{f44}) corresponds to the proper distance
\begin{equation}\label{f45}
 d_H(t;k)=a(t;k) \int_0^t \frac{dt'}{a(t';k)}   
\end{equation} 
when lengths are measured with $\langle g_{\mu\nu}\rangle_k$. Note also that with the
more general metric (\ref{cosmet1}) eq.~(\ref{f44}) gets replaced by
\begin{equation}\label{f46}
 r_H(t;k)=\int_0^t dt' \frac{b(t';k)}{a(t';k)}.   
\end{equation}
This indicates again that the second, undetermined metric function can acquire
a physical significance when RG effects are taken into account.

More information would probably be available when one investigates what 
properties a ``reasonable" state would have, particularly in the matter sector.
In late cosmology, where matter is given as a set of particles
with only weakly $k$-dependent masses $m(k)$,
the running of $\rho(t;k)$ essentially boils down to the running of the scale factor.
It is reasonable to assume that at any given time $t_1$, the number
of particles varies only very slightly with $k$, up to very high energies. Then
one can use information about the $k$-dependent masses of these particles 
to relate $\rho(t_1;k)$ with $a(t_1;k)$
(the number density of particles is proportional to $a(t_1;k)^{-3}$), now considered as
functions of $k$ only:
\begin{equation}
 \rho(t_1;k)=m(k)\, n \, a(t_1;k)^{-3},  
\end{equation}
where $n$ is the number density per unit coordinate volume. 
This determines the cosmological solutions to a greater extent.  

One of the main arguments used for the motivation of inflation is the so-called
``horizon problem". The statement is that at the time when the microwace background
radiation was emitted, far-separated regions of the universe had very similar properties
although they had up to then no time to ``communicate" with each other, in the
framework of standard cosmology without inflation.
We wish to emphasize here the possibility that the horizon problem might be solvable 
by the $k$-dependent causality structure of spacetime, without inflation.
For the high energy processes of the very early universe
(e.g. in the Planck era), it is very likely that 
the most appropriate description is obtained when one uses a $k$-microscope with 
very large $k$. For such a view on the early universe, it may well be that regions
appear causally connected that would be far outside each other's horizons within
the classical description, i.e. with $\left < g_{\mu\nu}\right > _{k_{\rm lab}}$.
Note in particular that the term ``Planck era" refers to the set of spacetime points
which are separated from the big bang by less than a Planck time {\it when time is 
defined via $\left < g_{\mu\nu}\right > _{k_{\rm lab}}$}.

A first encouraging result indicating that quantum gravity might solve the horizon
problem without inflation was found in \cite{cosmo1} in the context of RG improved
field equations. Their solution corresponding to the very early universe does not have
a particle horizon!\\ 

\noindent{\bf\large 6. The concept of an intrinsic scale}

We continue to consider homogeneous and isotropic cosmologies described by metrics
of the form (\ref{cosmet1}). Let us consider an arbitrary physical structure in the QEG
spacetime which has a comoving length $\Delta x$. A typical example would be the 
wavelength of some perturbation. Then the running metric $\langle g_{\mu\nu}\rangle_k$
associates to the fixed {\it coordinate} length $\Delta x$ the running {\it proper}
length
\begin{equation}\label{g1}
 L(t;k)=a(t;k)\,\Delta x.   
\end{equation}
What would be a sensible choice for the scale $k$ when one tries to observe the 
$\Delta x$-object by means of a ``$k$-microscope"? Following the discusion of the 
``intrinsic scale" in ref.~\cite{minang} a plausible choice seems to be the scale
$k \equiv k_{\rm in}(\Delta x,t)$ at which the resolving power of the microscope,
$\ell(k)$, equals precisely the yet to be determined proper length of the object:
\begin{equation}\label{g1a}
 L\bigg(t;k_{\rm in}(\Delta x,t)\bigg)=\ell\bigg(k_{\rm in}(\Delta x,t)\bigg).   
\end{equation}
If $\ell=\ell(k)$ and the scale factor $a(t;k)$ are known, this condition 
yields the following implicit equation for $k_{\rm in}(\Delta x,t)$:
\begin{equation}\label{g2}
 a\bigg(t;k_{\rm in}(\Delta x,t)\bigg)\,\Delta x= \ell\bigg(k_{\rm in}(\Delta x,t)\bigg).   
\end{equation}
If this equation has a unique solution it is natural to define the folowing length
$L_{\rm in}(t)$ as the ($t$-dependent, but $k$-independent) proper length ``intrinsic"
to the $\Delta x$-object:
\begin{equation}\label{g3}
 L_{\rm in}(t)\equiv L\bigg(t;k_{\rm in}(\Delta x,t)\bigg)
= a\bigg(t;k_{\rm in}(\Delta x,t)\bigg)\,\Delta x.  
\end{equation}
As in classical cosmology, we may refer to the ratio of the object's proper length and
comoving length as a scale factor. However, in the present case this ratio
\begin{equation}\label{g4}
 L_{\rm in}(t)/\Delta x= a\bigg(t;k_{\rm in}(\Delta x,t)\bigg)\equiv a_{\rm in}(t;\Delta x)   
\end{equation}
yields a scale factor $a_{\rm in}(t;\Delta x)$ which itself depends on $\Delta x$ and, in
a sense, is ``intrinsic" to the $\Delta x$-object. Stated differently, objects of
different comoving size are affected by the cosmological expansion in different ways;
each of them has its own scale factor,
\begin{equation}\label{g5}
 L_{\rm in}(t)= a_{\rm in}(t;\Delta x)\,\Delta x,   
\end{equation}
and its own $b_{\rm in}(t;\Delta x)\equiv b(t;k_{\rm in}(\Delta x,t))$. Though 
surprising at first sight, the physical mechanism behind this phenomenon is clear: 
Objects of different sizes are optimally described by taking the gravitational
couplings at different scales, and as a result the effective spacetime they determine
is different from object to object.

A similar discussion applies to temporal proper distances. Let us consider two
events $P_1$ and $P_2$ which have identical $x^i$-coordinates and $t$-coordinates 
differing by an amount $\Delta t>0$ with respect to the system of coordinates in which
(\ref{cosmet1}) is written. According to the metric $\langle ds^2\rangle_k$ the
proper time elapsed between the events is
\begin{equation}\label{g5e1}
 T(t;k)=b(t;k)\,\Delta t.   
\end{equation}
As in the spatial case, one can try to adjust $k\equiv k_{\rm in}(\Delta t,t)$
in such a way that the resulting proper time matches exactly the resolving power:
\begin{equation}\label{g5e2}
 b\bigg(t;k_{\rm in}(\Delta t,t)\bigg)\,\Delta t= \ell\bigg(k_{\rm in}(\Delta t,t)\bigg).   
\end{equation}
If this equation has a unique solution the natural definition for the proper time distance
``intrinsic" to the two events is 
\begin{equation}\label{g5e3}
 T_{\rm in}(t)\equiv T\bigg(t;k_{\rm in}(\Delta t,t)\bigg)\,\Delta t
=b\bigg(t;k_{\rm in}(\Delta t,t)\bigg)
 \,\Delta t.   
\end{equation}
It is derived from the explicitly $\Delta t$-dependent Robertson-Walker metric with \\
$a_{\rm in}(t;\Delta t)\equiv a(t;k_{\rm in}(\Delta t,t))$ and
$b_{\rm in}(t;\Delta t)\equiv b(t;k_{\rm in}(\Delta t,t))$.

We shall illustrate the points made above employing the COM definition of the
resolving power $\ell(k)$ which was explained in section 3. In particular we assume that
the WKB approximation is valid so that approximately $\ell(k)=\pi /k$. In this case we have
$k_{\rm in}(\Delta t,t) = \pi/L_{\rm in}(t)$ which, when inserted into (\ref{g3}), 
leads to an implicit equation directly for $L_{\rm in}(t)$:
\begin{equation}\label{g6}
 L_{\rm in}(t)=a\bigg(t;\pi/L_{\rm in}(t)\bigg)\,\Delta x.   
\end{equation} 
For intrinsic proper time intervals we use the same $\ell(k)$, whence
\begin{equation}\label{g6t}
 T_{\rm in}(t)=b\bigg(t;\pi/T_{\rm in}(t)\bigg)\,\Delta t.   
\end{equation}
Next we turn to various examples.\\

\noindent{\bf 6.1. Example: de~Sitter family with $k$-dependent Killing vectors}

Our first example is the de~Sitter type family of metrics (\ref{cosmet2}) with
(\ref{adesitfp}) which has some of its symmetries implemented in a $k$-dependent way.
It is based upon the fixed point running $\Lambda(k)=\lambda_* k^2$.
For this family, eq.~(\ref{g6}) reads
\begin{equation}\label{g7}
 L_{\rm in}(t)=A \,\Delta x \exp\left [ \sqrt{\frac{\lambda _*}{3}}\frac{\pi t}
 {L_{\rm in}(t)}\right ].   
\end{equation}
This functional equation can be solved in terms of the Lambert $W$-function 
\footnote{By definition \cite{corless}, the $W$-function satisfies 
$W(x)\exp[W(x)]=x$, and $W \equiv W_0$ denotes its real branch analytic at $x=0$.}
$W_0$:
\begin{equation}\label{g8}
 L_{\rm in}(t)=\frac{\alpha \, t}{W(\alpha t / A \Delta x)}.   
\end{equation}
Here $\alpha \equiv \pi \sqrt{\lambda _* /3}$, and the corresponding intrinsic
scale factor reads 
\begin{equation}\label{g9}
 a_{\rm in}(t;\Delta x)=\frac{\alpha \, t}{\Delta x \, W(\alpha t/A \Delta x)}.   
\end{equation}
We see that it depends explicitly on the size of the
object whose size is measured with the corresponding metric. 
While it seems absurd from the point of 
view of classical Riemannian geometry, this phenomenon is very natural 
from a quantum field theory perspective. The 
$\Delta x$-dependence of $a_{\rm in}$ simply reflects the fact that large objects 
``feel" the value of the gravitational parameters on other scales than small objects
do. Since $\Lambda(k)\propto k^2 \propto 1/L^2$ is the smaller the larger $L$ is, 
small objects will appear to expand faster than larger ones.

We can display this behavior analytically by specializing for objects which are much 
smaller than $\alpha t$. If $y \equiv \alpha t/A \Delta x\gg 1$ we may use the 
following asymptotic expansion of $W_0$ for $x \rightarrow \infty$:
\begin{equation}
  W_0(y)= \ln y -\ln \ln y+ \frac{\ln \ln y}{\ln y}+...  
\end{equation}
Retaining only the first term we get approximately 
\begin{equation}\label{g9a}
 a_{\rm in}(t;\Delta x)=\frac{\alpha \, t}{\Delta x \, \ln (\alpha t/ A \Delta x)}.   
\end{equation}
Remarkably, apart from a logarithmic correction, $a_{\rm in}$ is a {\it linear}
rather than exponential function of time (as it was classically): 
$a_{\rm in}\propto t/ \Delta x$. This slowing down of the expansion by quantum 
effects is easy to understand: as the universe expands and $L_{\rm in}(t)$ becomes larger, the 
object considered feels an ever decreasing cosmological constant. The combined effect
of the exponential expansion and the continuous reduction of $\Lambda(k)$ is the
linear expansion (\ref{g9a}).

Furthermore, we can look at objects of different size at a fixed time. Because we have 
approximately $a_{\rm in}\propto t/ \Delta x$ we see that large objects have
indeed a lower expansion rate than small ones.

So far we concentrated on the innermost ``core" of the QEG spacetime. When we move 
``outward", $\Lambda(k)$ leaves the NGFP regime. If it approaches a constant value
below some $k$, the spacetime will look like a classical de Sitter space macroscopically.
If this constant is zero, it will approach a standard smooth Minkowski space on
large distance scales.

It is quite intriguing that even Minkowski space, 
if we put it under a sufficiently strong microscope, might show a complicated pattern
of cosmological constant driven ``expansions" which we usually consider in macroscopic
cosmology only. One has to be careful in applying this cosmology-type picture,
however, since it provides only a local description in the domain accessible by the
highly symmetric Robertson-Walker metric. 
In fact, our simple model for the observation of spacetime by a ``microscope"
assumes that even in presence of the (back-reacting!) microscope the universe is homogeneous
and isotropic. In realistic experiments this will not be the case presumably so that one has
to deal with families $\langle g_{\mu\nu}\rangle_k$ with less or no symmetry which are
clearly much harder to come by.\\  

\noindent{\bf 6.2. Example: de~Sitter family with $k$-independent Killing vectors}

Our second example is the ``anomaly free" de~Sitter family with 10 scale independent
Killing vectors. As we found in eq.~(\ref{solcos2}), it corresponds to   
\begin{equation}\label{g10}
 a(t;k)=\sqrt{\frac{\Lambda(k_0)}{\Lambda(k)}} \exp (H_0 t), \quad\quad
 b(t;k)=\sqrt{\frac{\Lambda(k_0)}{\Lambda(k)}},    
\end{equation} 
where $H_0 =\sqrt{\Lambda(k_0)/3}$.
We analyze this family for a model trajectory of the type
\begin{equation}\label{g11}
 \Lambda(k)=B \, k^\omega, \quad\quad B>0,   
\end{equation}
with an arbitrary constant
exponent $\omega \geq 0$. Important special cases include $\omega=0$
(classical dS), $\omega=2$ (fixed point regime) and $\omega=4$ ($k^4$-regime). 
Therefore, with the abbreviation $\gamma \equiv \sqrt{B/\Lambda(k_0)}$,
\begin{equation}\label{g12}
 a(t;k)=\gamma^{-1} k^{-\omega/2} \exp(H_0 t), \quad b(t;k)=\gamma^{-1}k^{-\omega/2}   
\end{equation}
so that the self-consistency condition (\ref{g6}) assumes the form
\begin{equation}\label{g13}
 L_{\rm in}(t)=\gamma^{-1}\left ( \frac{L_{\rm in}(t)}{\pi}\right )^{\omega/2}
 \exp(H_0 t)\Delta x.   
\end{equation}
Its solution is easily found:
\begin{equation}\label{g14}
 L_{\rm in}(t)=\pi^{\omega/(\omega-2)}\left ( \frac{\gamma}{\Delta x}\right )^{2/(\omega-2)}
 \exp\left[ -\left (\frac{2}{\omega-2}\right ) H_0 t \right]. 
\end{equation}
The corresponding scale factor reads
\begin{equation}\label{g15}
 a_{\rm in}(t;\Delta x)=\beta
 \exp\left[ -\left (\frac{2}{\omega-2}\right ) H_0 t \right].   
\end{equation}
where we introduced
\begin{equation}\label{g15a}
 \beta \equiv \gamma^{2/(\omega-2)}
 \left ( \frac{\pi}{\Delta x}\right ) ^{\omega/(\omega-2)}.
\end{equation}
Together with $b_{\rm in}(t;\Delta x)=b(t;\pi/L_{\rm in}(t))$, given explicitly by
\begin{equation}\label{g16}
  b_{\rm in}(t;\Delta x)=\beta
 \exp\left[ -\left (\frac{\omega}{\omega-2}\right ) H_0 t \right],   
\end{equation}
it constitutes the Robertson-Walker metric ``intrinsic" to structures of comoving size
$\Delta x$:
\begin{equation}\label{g17}
 \langle ds^2\rangle_{\rm in}\equiv \langle ds^2\rangle_{k_{\rm in}}
 =-b_{\rm in}^2(t;\Delta x) \, dt^2+ a_{\rm in}^2(t;\Delta x)\, d {\bf x}^2. 
\end{equation}
The line element (\ref{g17}) reads in explicit form
\begin{equation}\label{g18}
 \langle ds^2 \rangle_{\rm in} = \beta^2 \exp \left [ \left ( \frac{2 \omega}{\omega -2} 
 \right ) H_0 t \right ] \left\{ -dt^2 + \exp(2H_0 t)d {\bf x}^2 \right\}.   
\end{equation}
As it should be, this equation coincides with $\langle ds^2 \rangle_{\rm in}
=[\Lambda(k_0)/\Lambda(k_{\rm in})]\langle ds^2 \rangle_{k_0}$. In fact, from (\ref{g14})
and $k_{\rm in}=\pi/L_{\rm in}$ we obtain the following expression for the 
intrinsic scale:
\begin{equation}\label{g19}
 k_{\rm in}(t;\Delta x)=\left ( \frac{\Delta x}{\pi\gamma}\right )^{2/(\omega-2)}
 \exp\left [ \left ( \frac{2}{\omega-2}\right ) H_0 t\right ].   
\end{equation}
It is instructive to rewrite it in terms of the proper length $L(t;k_0)$ measured 
with $\langle ds^2 \rangle_{k_0}$:
\begin{equation}\label{g20}
 k_{\rm in}(t;\Delta x)=k_0^{\omega/(\omega-2)}(L(t;k_0)/\pi)^{2/(\omega-2)}.   
\end{equation}

The above results are somewhat surprising and several comments are in order here.
For $\omega=0$ we recover the classical results: the scale factor grows
as $a_{\rm in}\propto\exp(H_0 t)$ and is independent of $\Delta x$, $b_{\rm in}$ 
is constant, and eq.~(\ref{g20}) yields the expected inverse proportionality
$k_{\rm in}\propto 1/L(t;k_0)$.

For a fixed $\Delta x$, we are free to bring the ``intrinsic" metric into the ``true"
Robertson-Walker form, i.e. to adjust the time coordinate such that $b_{\rm in}=1$
at all times. For $\omega\neq 0,2$ we define
\begin{equation}\label{g20a}
 \tilde{t}\equiv \int b_{\rm in}(t;\Delta x)dt = \beta \,\frac{2-\omega}{\omega H_0}
 \left ( \exp \left [\frac{\omega}{2-\omega}H_0 t \right ]-1\right ).  
\end{equation} 
We fixed the integration constant such that $\tilde{t}(t=0)=0$. Now obviously
$b_{\rm in}(\tilde{t},\Delta x)=1$, and the scale factor becomes
\begin{equation}\label{g20b}
 a_{\rm in}(\tilde{t};\Delta x)=\beta \left [ 1+\frac{\omega H_0 \tilde{t}}{(2-\omega)\beta}
 \right ]^{2/\omega}.   
\end{equation}
The cosmology with the scale factor (\ref{g20b}) is quite remarkable:
For $\omega >2$ it describes a contracting rather than expanding universe,
even though $a(t;k)\propto\exp(+H_0 t)$ grows for every fixed value of $k$. Moreover,
the ``intrinsic history" of the object of size $\Delta x$ ends in a ``big crunch" 
singularity at
\begin{equation}\label{g20c}
 \tilde{t}=\frac{(\omega-2)\beta}{\omega H_0}.   
\end{equation}

The case $\omega=4$ is particularly relevant since the middle part of most trajectories
is well approximated by $\Lambda\propto k^4$, $G=const$. For the separatrix this 
``$k^4$-regime" even extends to $k=0$. For $\omega =4$, the scale factor (\ref{g20b}) is
\begin{equation}\label{g20d}
 a_{\rm in}^{\omega=4}(\tilde{t},\Delta x)=\frac{\gamma\pi^2}{(\Delta{\bf x})^2}
 \sqrt{1-\frac{2 H_0 (\Delta{\bf x})^2}{\gamma\pi^2}\tilde{t}}.   
\end{equation}
Its ``big crunch" singularity is not reached actually, since $\Delta x$
at some point leaves the $\omega=4$-regime and enters the fixed point regime with
$\omega=2$.
Furthermore, eq.~(\ref{g20}) becomes
\begin{equation}\label{g22}
 k_{\rm in}^{\omega=4}(t;\Delta x)=k_0^2 L(t;k_0)/\pi,   
\end{equation}
i.e. rather than inversely, the scale $k_{\rm in}$ is {\it directly} proportional 
to $L(t;k_0)$.

These results seem to be counter-intuitive, but they have a natural physics explanation,
in fact. The effective time evolution of $a_{\rm in}$ is given by the equation
\begin{equation}\label{g23}
 \frac{d}{dt}a_{\rm in}(t;\Delta x)=
 \partial_t a(t;k_{\rm in}(t;\Delta x))+\bigg(\partial_t k_{\rm in}(t;\Delta x)\bigg)\,
 \partial_k a(t;k)|_{k=k_{\rm in}(t;\Delta x)}.  
\end{equation}
The first term on the right hand side of (\ref{g23}) 
is the time derivative at fixed $k$ which is dictated 
by the effective field equations at the fixed scale $k$. In our case it is given by the 
exponential growth $\propto \exp(H_0 t)$.
The second term accounts
for the rescaling of $a_{\rm in}$ due to the change of $k$. 
It is in general cosmologies a property of the particular
state considered and is usually only to a small extent determined by the flow 
equations. In our highly symmetric case it is determined by the factor
$\sqrt{\Lambda(k_0)/\Lambda(k)}$. The term $\partial_t k_{\rm in}(t;\Delta x)$ is itself
a function of the LHS, $(d/dt) a_{\rm in}$, and therefore (\ref{g23}) is an implicit
equation.
For $\omega>0$, the term $\partial_k a(t;k)$ amounts to a ``shrinking" of space
with increasing $k$. The effective contraction described by (\ref{g20b}) is explained
by noting that for $\omega>2$ this shrinking overcomes the exponential growth
from the first term, with the result of a decreasing $a_{\rm in}$.  

The explanation for eq.~(\ref{g22}) is similar. The shrinking due to the large negative
value of $\partial_k a(t;k)$ is so strong that a larger coordinate distance
$\Delta x \equiv L(t;k_0)/a(t;k_0)$ corresponds to a smaller proper length
$L_{\rm in}\equiv \Delta x a(t;k_{\rm in})$ and a larger value of $k_{\rm in}$.
This is completely analogous to the shrinking $S^4$ discussed in \cite{minang}.
See also the more detailed discussion there.\\

Finally we note that the notion of an ``intrinsic length" breaks down 
directly at the non-Gaussian fixed point, i.e. when $\omega=2$ exactly.
This can be already seen from eq.~(\ref{g13}). In the fixed point regime, (\ref{g13})
becomes
\begin{equation}\label{ill}
 L_{\rm in}(t)=\Delta x \frac{L_{\rm in}(t)}{\pi}
 \sqrt{\frac{\Lambda(k_0)}{\lambda_*}}\exp (H_0 t).    
\end{equation}
$L_{\rm in}(t)$ appears on both sides linearly and drops out, 
so the equation cannot be solved.
We already met this ill-defined situation in our previous paper \cite{minang} in the 
case of the four-sphere, which had a $k$-dependent radius which was also proportional to
$\Lambda(k)^{-1/2}$. In the fixed point regime, there is only one value of $\Delta x$
which can be observed with an ``appropriate" microscope. Within our parametrization the
value is time dependent,
\begin{equation}\label{soldx}
 \Delta x=\pi \sqrt{\frac{\lambda _*}{\Lambda(k_0)}}\exp(-H_0 t).   
\end{equation} 
When one tries to zoom deeper into this coordinate separation, spacetime locally ``shrinks"
in a way so that it precisely cancels the effect of zooming. 

To explain this, we note 
that in the fixed point regime $L(t;k)$
is proportional to $1/k$. This means that spacetime ``shrinks" when we increase the 
resolution $k$. The length $L_{\rm in}(t)$ 
usually arises by finding an ``appropriate" value of $k$,
$k=k_{\rm in}(t;\Delta x)$, so that $L_{\rm in}$ fulfills 
eq.~(\ref{g1a}). In 
the example of section 6.1. we found for any $t$ precisely one appropriate 
value of $k_{\rm in}$, as one would usually expect. 
But now, with the relation $L(t;k)\propto 1/k$,
the correspondence breaks down. For the unique time at which eq.~(\ref{soldx}) is fulfilled
for some given $\Delta x$, {\it every} $k$ in the fixed point regime is an appropriate
value for $k_{\rm in}$, and therefore there is no preferred value for $L_{\rm in}$.
At any other time, in contrast, there is {\it no} solution
for $k_{\rm in}$ and therefore again no preferred value for $L_{\rm in}$.

These considerations show that the approximate and
somewhat heuristic notion of an intrinsic scale of an object
and the spacetime geometry ``felt" by this object does not always work as
unambiguously as in the example of section 6.1.\\

One obtains similar results for intrinsic proper distances in the time direction.
The scale is $t$-independent in this case,
\begin{equation}\label{g30}
 k_{\rm in}(t;\Delta t)=\left ( \frac{\Delta t}{\pi \gamma}\right )^{2/(\omega-2)},   
\end{equation}
and the metric coefficients read
\begin{equation}\label{g31}
 a_{\rm in}(t;\Delta t)=\gamma ^{2/(\omega-2)}\left ( \frac{\pi}{\Delta t}
 \right )^{\omega/(\omega-2)}\exp(H_0 t),   
\end{equation}
\begin{equation}\label{g32}
  b_{\rm in}(t;\Delta t)=\gamma ^{2/(\omega-2)}\left ( \frac{\pi}{\Delta t}
 \right )^{\omega/(\omega-2)}.   
\end{equation}
Here $a_{\rm in}$ has the classical time dependence $\propto \exp(H_0 t)$
for any value of $\omega$, and $b_{\rm in}$ is time independent. The relationship between
$\Delta t$ and the intrinsic proper time interval is likewise $t$-independent:
\begin{equation}\label{g33}
 T_{\rm in}=\pi^{\omega/(\omega-2)}\left ( \frac{\gamma}{\Delta t}
 \right )^{2/(\omega-2)}.   
\end{equation} 
If $\omega=4$, for example,
\begin{equation}\label{g34}
 T_{\rm in}^{\omega=4}=\pi^2 \gamma \frac{1}{\Delta t}.   
\end{equation}
The ``shrinking" of spacetime with growing $k$ is again so strong that a
{\it larger} coordinate interval $\Delta t$ corresponds to a {\it smaller}
proper intrinsic time $T_{\rm in}$. \\

\noindent{\bf 6.3. Example: Robertson-Walker cosmology with relativistic fluid}

In order to connect our formalism to results obtained in earlier work
(``RG improved field equations" \cite{cosmo1}), 
we finally consider a Robertson-Walker spacetime
filled by a relativistic fluid.
We restrict ourselves to the following situation:
\begin{itemize}
\item $b(t;k)=1$ for all $t$ and $k$.
\item The equation of state parameter $w$ is $k$-independently $1/3$, so that the 
solution eq.~(\ref{f37}) is valid for every $k$.
\item The parameter $\mathcal{M}$ introduced in eq.~(\ref{f37}) is $k$-independent.
\item We consider the fixed point running of $G$ and $\Lambda$, eq.(\ref{f39}).
\item $k \ll 1/t$ or $k \gg 1/t$, so that either eq.~(\ref{f41}) or eq.~(\ref{f42})
is valid.  
\end{itemize}
The motivation for these specializations (except the last) is that the RG improvement
discussed in \cite{cosmo1} applies to the same situation. 
Starting with eq.~(\ref{f41}), we have to solve
\begin{equation}\label{rw1}
 L_{\rm in}(t)=\Delta x \, a \left ( t,k=\frac{\pi}{L_{\rm in}}\right )
 = \Delta x \, (\alpha\mathcal{M})^{1/4} \sqrt{t L_{\rm in}/\pi}   
\end{equation}
(where $\alpha\equiv g_*/6$) which yields the linear growth 
\begin{equation}\label{rw2}
 L_{\rm in}(t)= (\Delta x)^2 (\alpha\mathcal{M})^{1/2}\, \frac{t}{\pi}.   
\end{equation}
The condition for the validity
of eq.~(\ref{f41}), $k_{\rm in}\ll 1/t$, implies
\begin{equation}\label{rw3}
t \ll \frac{L_{\rm in}}{\pi}=\left ( \frac{\Delta x}{\pi}\right ) ^2 
(\alpha\mathcal{M})^{1/2} t   
\end{equation}
or
\begin{equation}\label{rw4}
 \Delta x \gg \frac{\pi}{(\alpha\mathcal{M})^{1/4}}.   
\end{equation}
In the universe we live in the parameter $\mathcal{M}$ relevant to the radiation
dominated epoch is of the order of magnitude
\begin{equation}\label{rw5}
 \mathcal{M}^{1/4}\approx 10^{-30} \frac{a_0}{\ell_{\rm Pl}},   
\end{equation}
where $a_0$ is the value of the scale factor today. Assuming that $g_*$ is of the order 1
we obtain that the above approximation is valid for objects which have today 
a size
\begin{equation}\label{rw6}
 L_{\rm today}=\Delta x a_0 \gg 10^{30}\ell_{\rm Pl}\approx 10^{-3}{\rm cm}.   
\end{equation}

The same analysis for eq.~(\ref{f42}) amounts to solving the equation
\begin{equation}\label{rw7}
 L_{\rm in}(t)=\Delta x \, a \left ( t,k=\frac{\pi}{L_{\rm in}}\right )
 = \Delta x \,\frac{L_{\rm in}}{\pi} (\beta\mathcal{M})^{1/4}
 \exp \left ( \sqrt{\frac{\lambda_*}{3}}\frac{\pi}{L_{\rm in}}\, t \right ),
\end{equation}
where now $\beta\equiv g_*/2 \lambda_*$. The solution for $L_{\rm in}$ is
again linear in $t$:
\begin{equation}\label{rw8}
 L_{\rm in}(t)=\pi t \, \sqrt{\frac{\lambda_*}{3}}
 \left\{ \log \left[ \frac{\pi}{\Delta x}\left ( \frac{1}{\beta\mathcal{M}}
 \right )^{1/4}\right]\right\} ^{-1}.   
\end{equation}
The condition for the validity
of eq.~(\ref{f42}), $k_{\rm in}\gg 1/t$, implies
\begin{equation}\label{rw9}
 t \gg \frac{L_{\rm in}}{\pi}= t \, \sqrt{\frac{\lambda_*}{3}}
 \left\{ \log \left[ \frac{\pi}{\Delta x}\left ( \frac{1}{\beta\mathcal{M}}
 \right )^{1/4}\right]\right\} ^{-1}
\end{equation}
or
\begin{equation}\label{rw10}
 \Delta x \ll \frac{\pi}{(\beta\mathcal{M})^{1/4}}
 \exp \left ( -\sqrt{\frac{\lambda_*}{3}}\right ).   
\end{equation}
Assuming that $\beta$ and $\exp \sqrt{\lambda_*/3}$ are of order 1, we get for 
the case of our universe that the above analysis is valid if
\begin{equation}\label{rw11}
 L_{\rm today}=\Delta x a_0 \ll 10^{30}\ell_{\rm Pl}\approx 10^{-3}{\rm cm}.   
\end{equation}

In both cases, $k \ll 1/t$ and $k \gg 1/t$, 
we found a linear intrinsic expansion $a_{\rm in}\propto t$, valid as long
as the object is well inside the fixed point regime. 
In the interpolating transition region $k \approx 1/t$ we expect a qualitatively 
similar behavior. A universe
with such a time dependence of the scale factor has no 
particle horizon according to eq.~(\ref{f44}). So we have indeed found a case in which
the horizon problem does not occur and therefore cannot serve as an argument
for inflation. The result $a_{\rm in}\propto t$ obtained here gives independent
support to the linear expansion found by RG improving the field equations 
\cite{cosmo1}.\\

\noindent{\bf\large 7. Summary}

In this paper we analyzed various conceptual issues related to a scale dependence 
of the metric. The discussion is relevant to the asymptotic safety scenario for
gravity and, more generally, to the analysis of all phenomena with a strong RG running
of the average metric. We described the role of the running effective field equations
implied by the average action of QEG and their solutions $\{ \langle g_{\mu\nu}\rangle_k
, 0\leq k < \infty \}$. 

The field equations derived from the effective average actions $\{ \Gamma_k \}$
allow for infinitely many solutions at each value of $k$. We can only determine from them
the set of solutions at any separate value of $k$, but not the evolution
$k \mapsto \langle g_{\mu\nu}\rangle_k $ corresponding to a particular quantum 
state $| \Psi \rangle$. 
We observed two sources of ambiguities which cannot
be resolved without knowledge of $| \Psi \rangle$. First, integration constants
like the parameter $M$ in the classical Schwarzschild solution become  
functions of $k$ in the quantum case. Second, simplifications of the metric due
to appropriate coordinate transformations can be made for one value of $k$ only.
An example was the Robertson-Walker metric
$ds^2= -b(t;k)^2 dt^2 + a(t;k)^2 d {\bf x}^2$. While $b$ can be set equal to one
by a redefinition of the time coordinate in classical gravity, this is possible for
only one chosen value of $k$ in the quantum case. We have therefore two functions
of $t$ and $k$ instead of one, 
but the field equations still determine only one of them.

Only for a maximally symmetric spacetime, i.e. de Sitter space, it was possible to
determine the evolution $k \mapsto \langle g_{\mu\nu}\rangle_k $
completely from the field equations and special symmetry requirements.
We found that the de Sitter metric scales as 
$\langle g_{\mu\nu}\rangle_k \propto \Lambda(k)^{-1}$.

The scale dependent metric as well as the scale dependent structure of the 
propagators lead to a scale dependent notion of causality. 
Outside the classical regime, the position or even
existence of horizons generically depend on the field chosen to transmit a signal
and on the value of $k$ relevant for the transmission process.    

This is particularly interesting for the early universe, since it might surround the
necessity of inflation: The particle horizons leading to the so-called horizon problem
(which is one of the main arguments for inflation) are the {\it classical} horizons
which may be irrelevant at typical scales governing processes 
in the very early universe.

One of the central themes of this paper is the different status enjoyed by
$k$-dependent and $k$-independent diffeomorphisms. We saw that the group of gauge
transformations consists of the $k$-independent ones only and that this is
one of the reasons why the effective field equations cannot completely determine
the gauge invariant, i.e. ``physical" contents of the family 
$\{ \langle g_{\mu\nu}\rangle_k \}$. Depending on whether Killing vectors 
are $k$-independent or not they either implement a symmetry on the manifold of 
physical events or they are ``anomalous".

We also discussed the possibility of assigning an intrinsic length
to objects living in a QEG spacetime, defined as the 
proper length of an object when observed by a ``microscope" which can just resolve it,
and we investigated under which conditions this can be a meaningful notion.\\   

\noindent{\bf Acknowledgements}

M.~R. would like to thank M.~Niedermaier and R.~Percacci for helpful discussions. 
He is also grateful to the Albert Einstein Institute for the hospitality extended
to him while this work was in progress.

\end{document}